\documentclass{aa}

\usepackage{graphicx}
\usepackage[svgnames]{xcolor}
\usepackage[pdftex,colorlinks,citecolor=DarkBlue]{hyperref}

\usepackage[varg]{txfonts}

\usepackage{natbib}
\bibpunct{(}{)}{;}{a}{}{,} 

\titlerunning{From intermediate galactic scales to self-gravitating cores}
\authorrunning{Hennebelle}

\begin{document}

\title{The FRIGG project: From intermediate galactic scales to self-gravitating cores}

\author{Patrick Hennebelle\inst{1,2}}
\institute{Universit\'e Paris Diderot, AIM, Sorbonne Paris Cit\'ee, CEA, CNRS, F-91191 Gif-sur-Yvette, France. \\
\and
LERMA (UMR CNRS 8112), Ecole Normale Sup\'erieure, 75231 Paris Cedex, France
}

\abstract{Understanding the detailed structure of the interstellar gas is essential for 
 our knowledge of the star formation process.} 
{The small-scale structure of the interstellar medium (ISM) is a direct consequence of the  galactic scales and 
making the link between the two is essential.} 
{We perform adaptive mesh simulations that aim to bridge the gap between the intermediate 
galactic scales and the self-gravitating prestellar cores. For this purpose we use stratified 
supernova regulated ISM magneto-hydrodynamical (MHD) simulations at the kpc scale to set up the initial 
conditions. We then zoom, performing a 
series of concentric uniform refinement and then refining on the Jeans length for the last levels. 
This allows us to reach a spatial resolution of a few $10^{-3}$ pc. 
The cores are identified using a clump finder and various criteria based on virial analysis. 
Their most relevant properties are computed and, due to the large number of objects formed in 
the simulations, reliable statistics are obtained. }
{The cores properties show encouraging 
agreements with observations.
The mass spectrum presents a clear powerlaw at high masses with an 
exponent close to $\simeq -1.3$ and a peak at about 1-2 $M_\odot$. 
The velocity dispersion and the angular momentum
distributions are respectively a few times the local sound speed and a few $10^{-2}$ pc km s$^{-1}$.
We  also find that the distribution of thermally supercritical cores  present a  range of 
magnetic mass-to-flux over critical mass-to-flux  ratio which typically ranges between $\simeq$0.3 and 3.
 indicating that they are significantly magnetized.
Investigating the time and spatial dependence of these statistical properties, we 
conclude that they are not significantly affected by the zooming procedure and that they do not 
present very large fluctuations. The most severe issue appears to be the dependence on the 
numerical resolution of the CMF. While the core definition process may possibly introduce some biases, 
the peak tends to shift to smaller values when the resolution improves. } 
{Our simulations, which use self-consistently generated initial conditions at the kpc scale, produce a large number of prestellar 
cores from which 
reliable statistics can be inferred. Preliminary comparisons with observations show encouraging agreements.  In particular 
the inferred CMF resemble the ones inferred from recent observations. We stress, however, a possible issue with
the peak position shifting with numerical resolution.}


\maketitle

\section{Introduction}

One the most difficult aspect which limits our understanding of the star 
formation process is its multi-scale nature. While the conditions
which lead to the formation of molecular clouds, where star birth takes place, 
are induced by the large and intermediate galactic scales, the ultimate 
mass reservoir of stars, the prestellar cores, are only a few 10$^{-2}$ pc 
wide \citep[e.g.][]{ward2007,offner2014}. This implies that ideally one would need to get a continuous 
description of spatial scales going from at least a few hundreds  to a few hundredths of pc.

Various studies have investigated the core formation in simulations and 
fewer have attempted to provide statistics of the core properties. 
Typically a box of a few pc across is specified with a prescribed mean density and 
velocity dispersion and the turbulence is either driven or free to decay.
The simulations are either hydrodynamical or  magneto-hydrodynamical 
\citep{klessen1998,klessen2001,klessen2005,vazquez2005,offner2008,dib2010,gong2011,gong2015}
and some model  the ambipolar diffusion \citep{vanloo2008,kudoh2008,kudoh2011,chen2014}.
While this kind of approach offers a natural framework to study  the core
formation in details, they encounter two difficulties. First of all, the core properties, 
such as their mass distribution directly depends on the simulation parameters \citep[e.g.][]{klessen2000,gong2015}
such as the mean Jeans mass, 
thus it is necessary to perform an ensemble of simulations and for the purpose of comparing 
with observations, to convolve the core distribution by the distribution of large scale initial 
conditions. Second the number of cores formed is often restricted to a small number
limiting the inferred statistics.

In an attempt to circumvent these two difficulties but also to 
bridge the gap between the intermediate galactic scales, that is to 
say the scales of a few hundreds of pc and the scales of the self-gravitating prestellar 
cores, that is on the order of 0.1 pc, we perform zooming simulations starting from 
self-consistently generated initial conditions. 
The benefit of this approach is that there is no need to specify 
the initial conditions of the dense molecular phase. A distribution of molecular 
clouds is naturally produced from the diffuse atomic gas. 
The initial setup is 
very similar to the studies  described in 
\citet{hennebelle2014} and \citet{iffrig2017}
\citep[see also][]{korpi+1999,slyz+2005,deavillez+2005,Joung06,Hill12, kim+2011, kim+2013, gent+2013,gatto2015}.
These studies consider a kpc stratified galactic box. The ISM is self-regulated 
by the star formation process and the associated supernova explosions, which inject 
energy and momentum and sustain the turbulence.
The finest spatial resolution obtained in the simulations is a few 10$^{-3}$ pc and  allows us to describe the formation 
of cores of mass larger than a few 0.1 $M_\odot$ while the hundred of pc size region where full zooming is 
applied leads to a large number of cores from which reliable statistics can be obtained. 

Note that other zooming simulations have been performed in the context of the star formation studies, such
as
for example the ones by \citet{offner2008} and \citet{padoan2014},
 which  started from molecular cloud scales and zoom up to few tens of AU ones.
At the kpc scale, the deepest zoom simulations have been performed by \citet{butler2015}, where the spatial resolution
goes up to 0.1 pc \citep[see also][]{seifried2017}  
To our knowledge the one presented here is the first to make the link  between few hundreds of pc  and few thousands of AU scales.

The plan of the paper is as follows.
In the second section, we describe the numerical setup, the physics included in the simulations
as well as the zooming procedure that we employed. The third section explains the 
algorithm used to identify the cores in 3D space and gives the definition of the computed quantities.
In the fourth section we present the structure and the core statistics obtained for various 
definition and criteria. In the fifth section, we look at various subregions and subset 
of cores to explore their dependence to environments. In section sixth, we investigate 
the time dependence of the statistics with the aim of assessing the robustness of the results. We also
compare the results obtained with three different spatial resolutions. Finally section seven concludes the paper.

\section{General setup}

\subsection{Code and processes}
To perform our simulations, we employ the code RAMSES
\citep{Teyssier02,Fromang06}, which is an adaptive mesh refinement 
code working in Cartesian geometry and using finite 
volume methods and Godunov solvers to solve the MHD equations. Ramses uses a constraint transport 
scheme for the magnetic field, which preserves $\rm{div} B$
to machine precision. 

As described below, we make an intense usage of the AMR scheme and 
starting from level 9 we introduce another eight to ten AMR levels, therefore reaching level 17-19.

      The simulations include various physical processes known to be important
      in the ISM. The ideal MHD
      equations with self-gravity are solved and take into account the cooling and heating
      processes relevant to the ISM, which 
       include UV heating and a cooling function with
      the same low-temperature part as in \citet{Audit05} and the
      high-temperature part based on \citet{Sutherland93}, resulting in a
      function similar to the one used in \citet{Joung06}. 

      An analytical gravity profile
      accounting for the distribution of stars and dark matter is added. The
      corresponding gravitational potential is given by \citep{Kuijken89b}:
      \begin{equation}
          \phi_{ext}(z) = a_1 \left(\sqrt{z^2 + z_0^2} - z_0\right) + a_2 \frac{z^2}{2},%
          \label{eq:ext-potential}
      \end{equation}
      with $a_1 = 1.42 \times 10^{-3}\ \mathrm{kpc}\ \mathrm{Myr}^{-2}$, $a_2 = 5.49 \times
      10^{-4}\ \mathrm{Myr}^{-2}$ and $z_0 = 180\ \mathrm{pc}$, as used by \citet{Joung06}.
     The gravitational
      potential $\Phi$ has thus two terms  the one due to stars and
      dark matter $\phi_{ext}$, and the one due to the gas itself $\phi$, hence
      $\Phi = \phi + \phi_{ext}$.

\subsection{Initial conditions}

      We initialize our simulations with a stratified disc: we use a Gaussian
      density profile:
      \begin{equation}
      n(z) = n_0 \exp \left[ - \frac{1}{2} \left( \frac{z}{z_0}  \right)^2 \right],
      \end{equation}
      where $n_0 = 1.5\ \mathrm{cm^{-3}}$ and $z_0 = 150\ \mathrm{pc}$. 
      This leads to a total column density, $\Sigma$, through the disc that is equal to 
      $\sqrt{2 \pi} \rho_0 z_0 $ where  $\rho_0=m_p n_0$ and $m_p= 2.3 \times 10^{-24}$ g is the mean mass per particle, 
which corresponds to a mixture of hydrogen and about 10 $\%$ of helium as in the ISM. 
      We get $\Sigma = 4 \times 10^{-3}$ g cm$^{-2} = 19.1$ M$_\odot$ pc$^{-2}$.


      The
      temperature is set to an usual WNM temperature, around $8000\ \mathrm{K}$.
      In order to prevent this disc from collapsing, an initial turbulent
      velocity field is generated with a RMS dispersion of $5\ \mathrm{km / s}$
      and a Kolmogorov \citep{Kolmogorov41} power spectrum with random phase.
      The initial horizontal magnetic field  is given by
      \begin{equation}
      B_x(z) = B_0 \exp \left[ - \frac{1}{2} \left( \frac{z}{z_0} \right)^2 \right],
      \end{equation}
      with $B_0 \simeq 3 \mathrm{\mu G}$.

\setlength{\unitlength}{1cm}
\begin{figure*} 
\begin{picture} (0,24.5)
\put(0,18){\includegraphics[width=8.7cm]{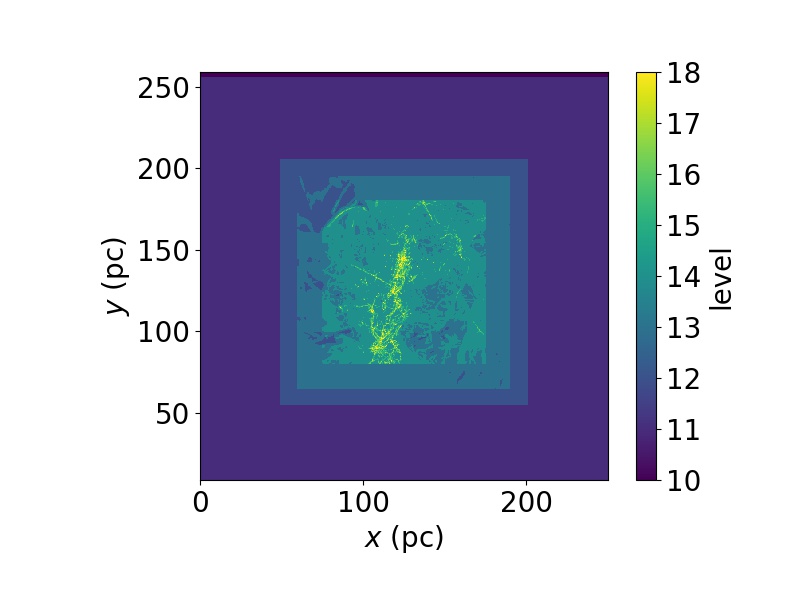}}  
\put(9,18){\includegraphics[width=8.7cm]{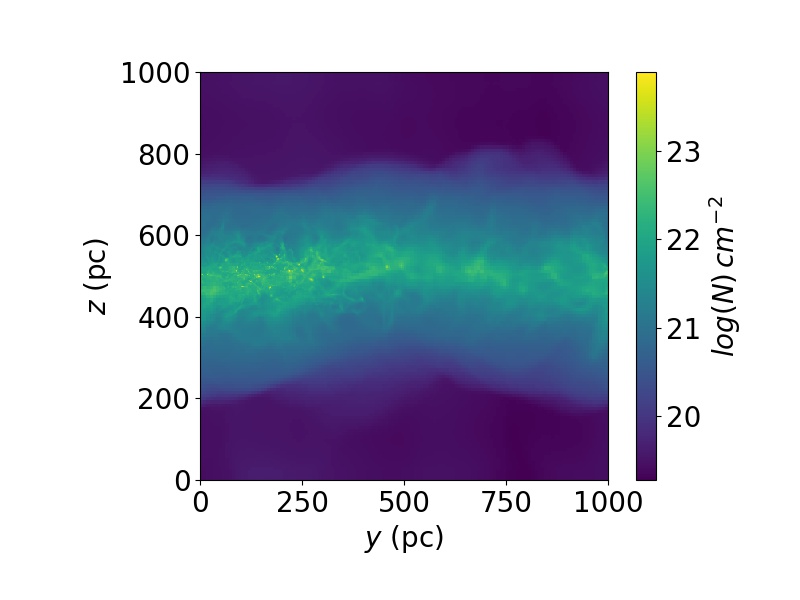}}  
\put(0,12){\includegraphics[width=8.7cm]{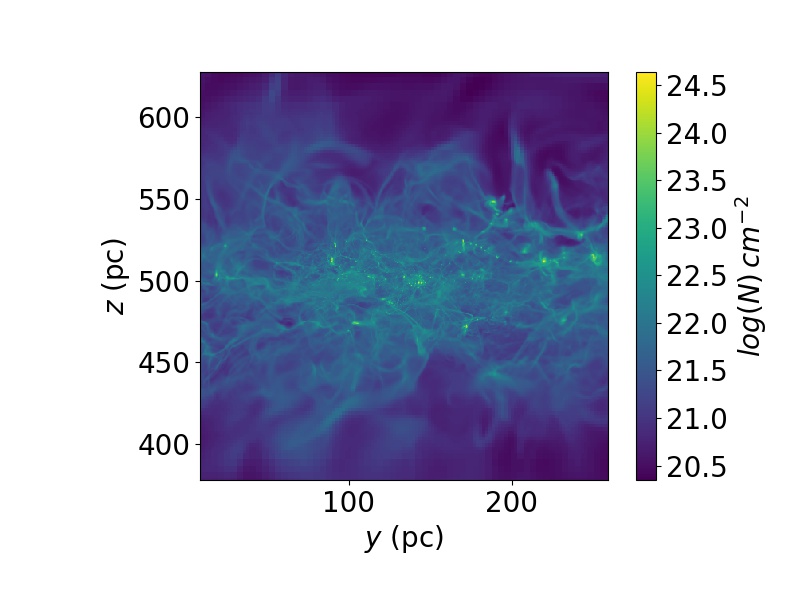}}  
\put(9,12){\includegraphics[width=8.7cm]{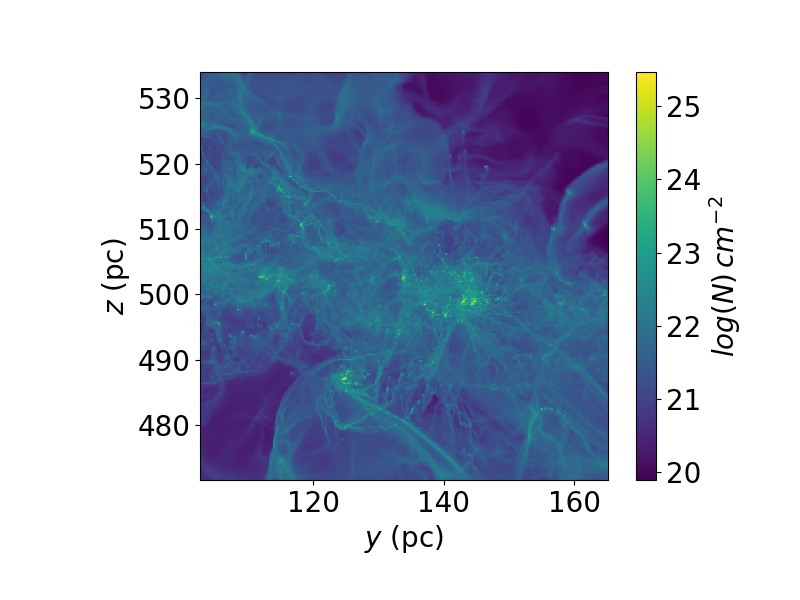}}  
\put(0,6){\includegraphics[width=8.7cm]{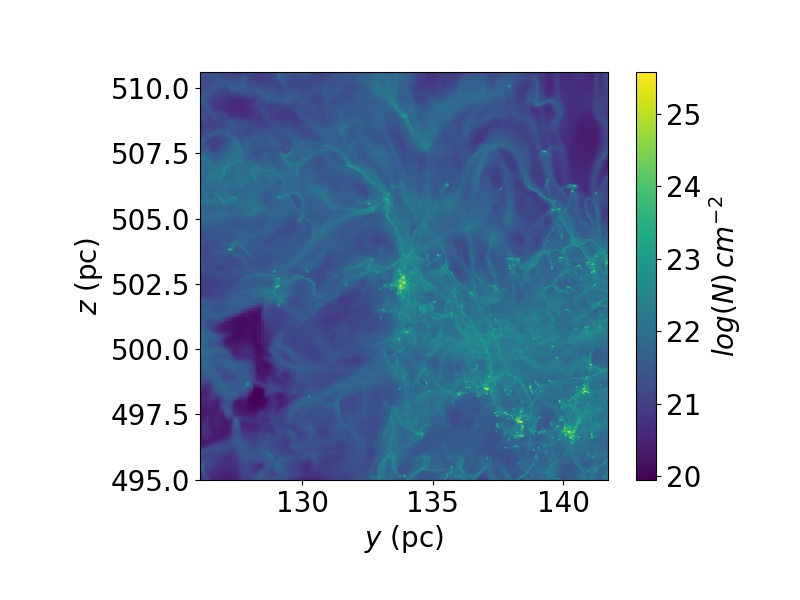}}  
\put(9,6){\includegraphics[width=8.7cm]{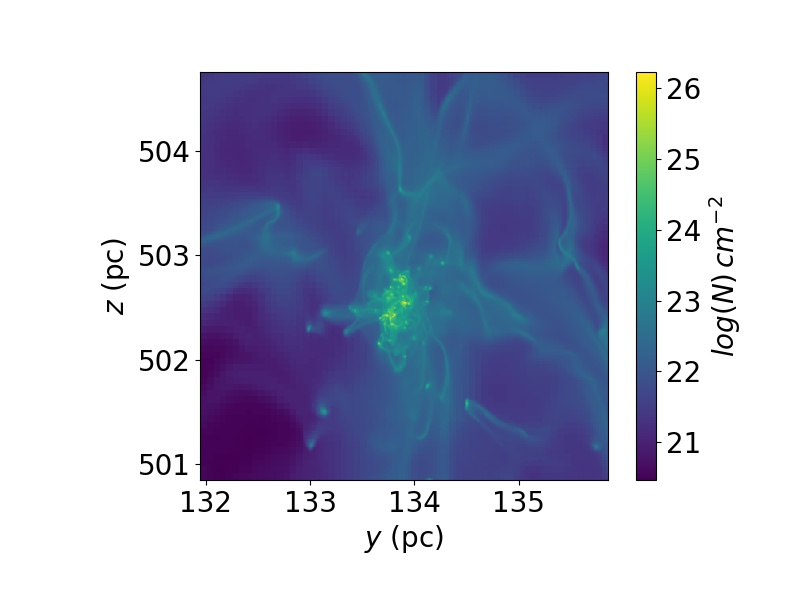}}  
\put(0,0){\includegraphics[width=8.7cm]{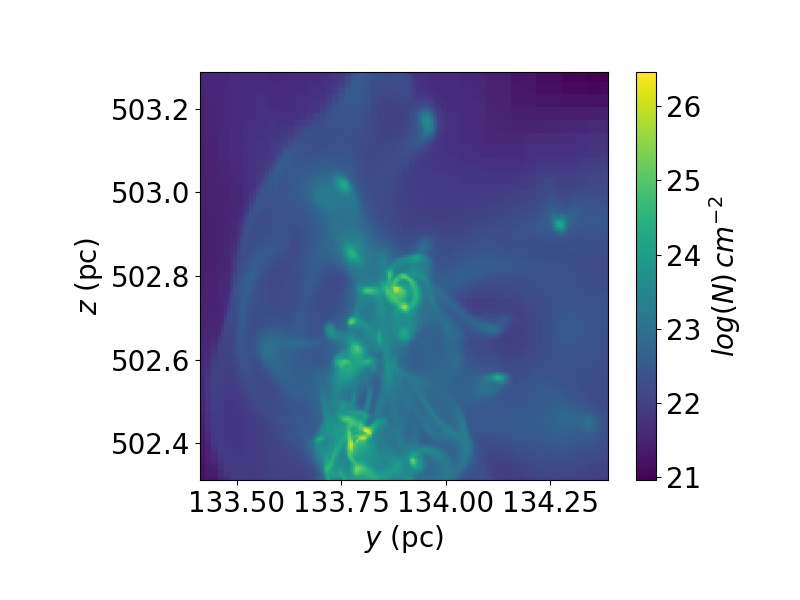}}  
\put(9,0){\includegraphics[width=8.7cm]{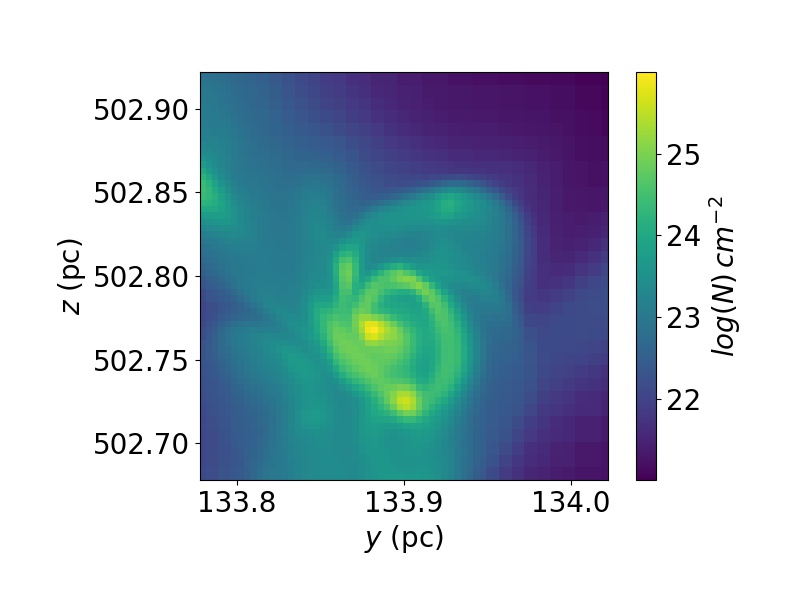}}  
\end{picture}
\caption{Top-left panel shows the AMR level  used to perform the calculation in one quarter of the computing box.
The zooming strategy is clearly visible. The first levels use nearly uniform refinement while the 
last ones are based on the Jeans length and follow the dense gas. 
Top-right panel shows the column density for the whole computing box and along the x-axis.  
Second, third and fourth rows display a  
series of zooms, going from 250 pc to 0.25 pc, showing  the column density along the y-axis.
From the bottom rows, the interest and limit of the calculation clearly appear. The cores as entities are 
reasonably described but their internal structure is poorly described.}
\label{zoom_series}
\end{figure*}

\setlength{\unitlength}{1cm}
\begin{figure*} 
\begin{picture} (0,11.5)
\put(8,5.5){\includegraphics[width=8cm]{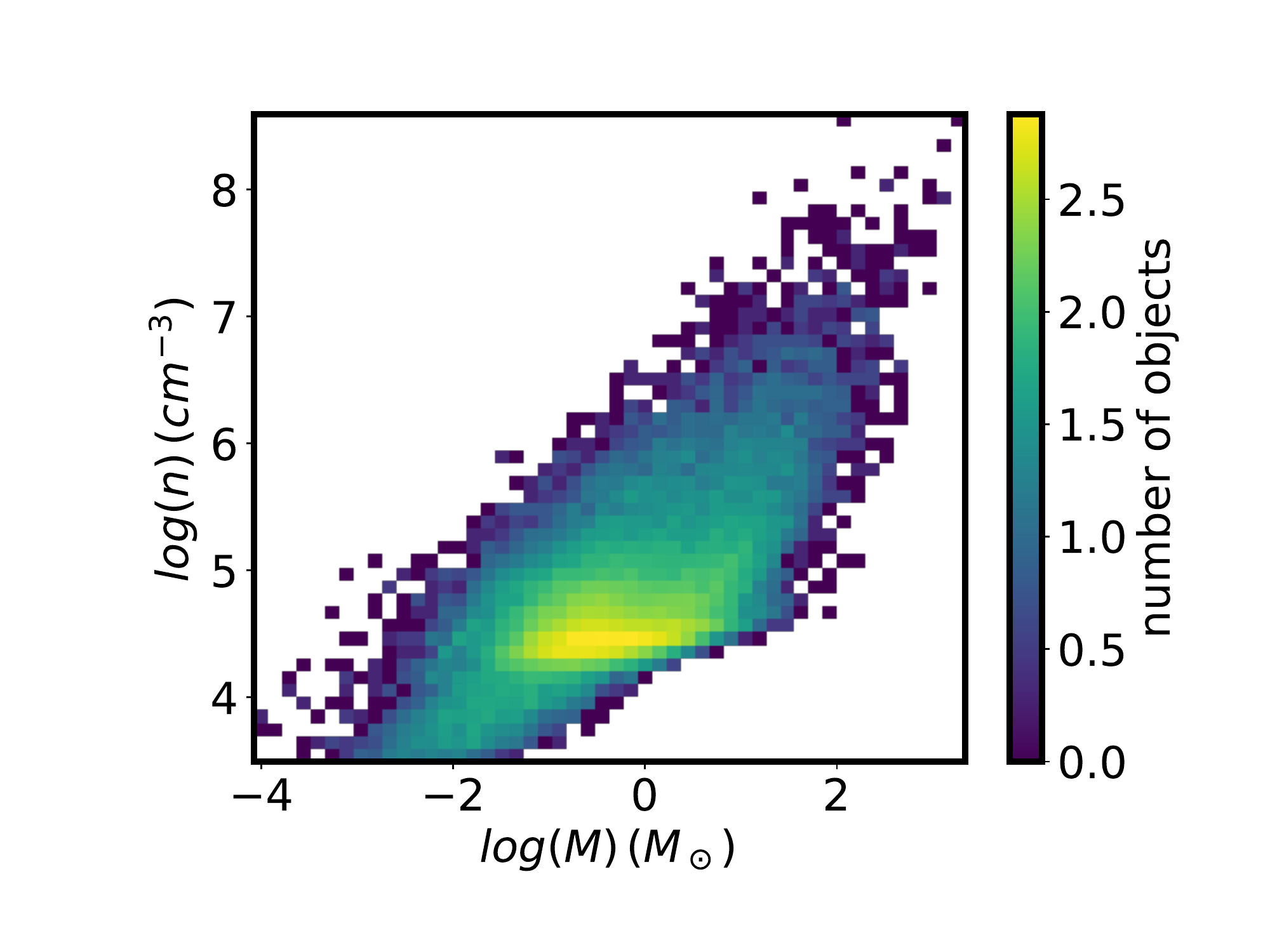}}  
\put(0,5.5){\includegraphics[width=8cm]{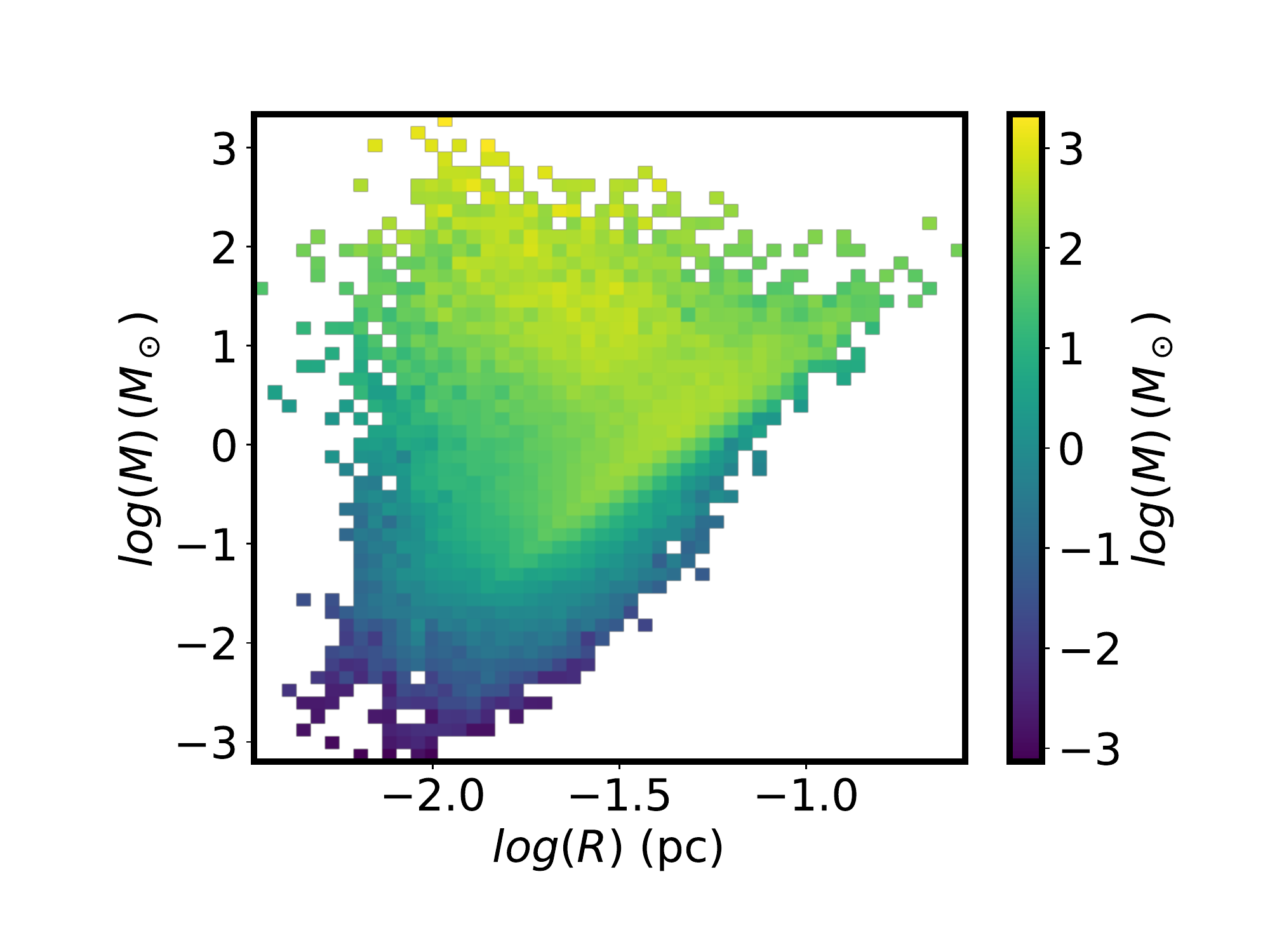}}  
\put(0,0){\includegraphics[width=8cm]{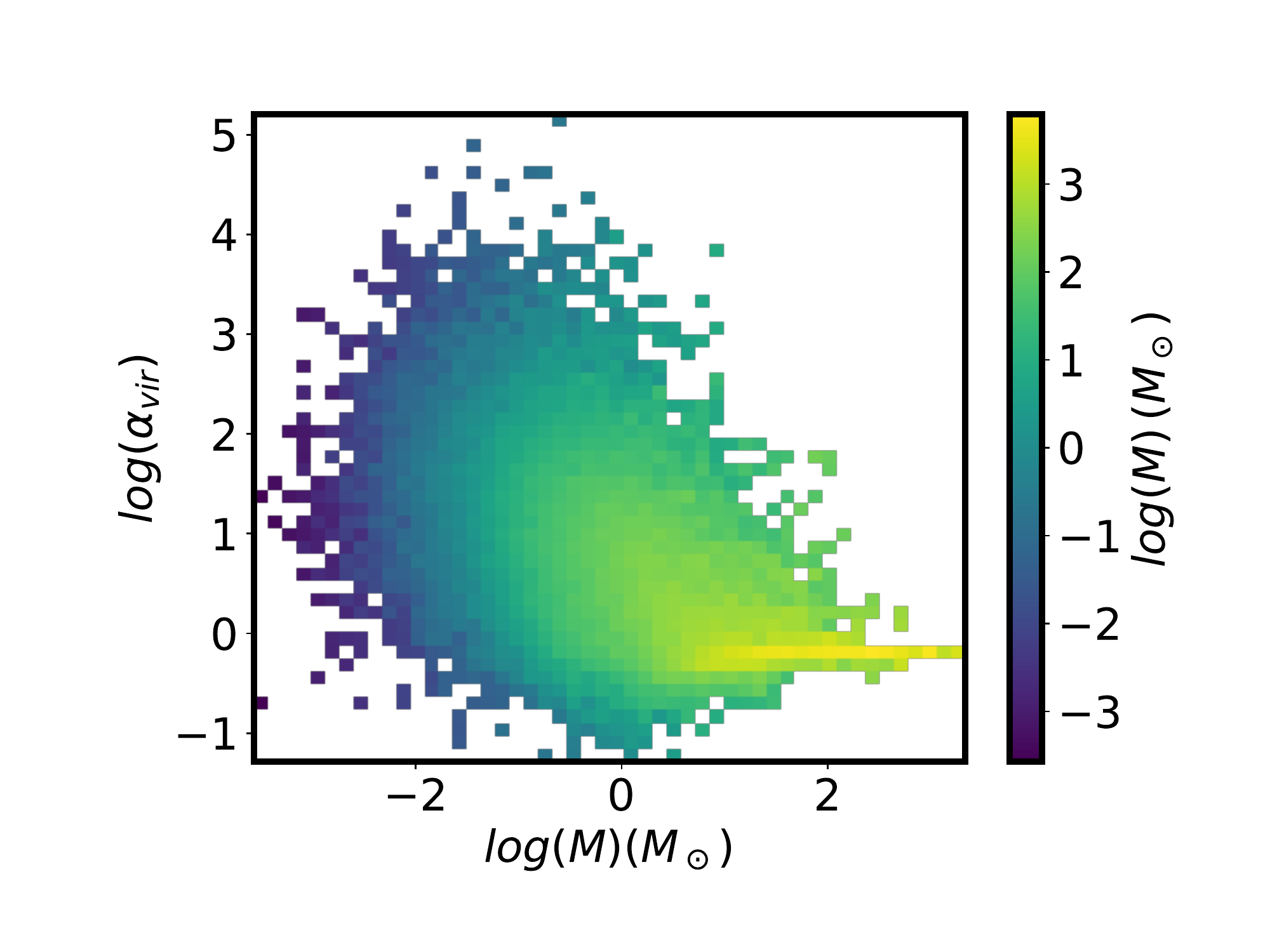}}  
\put(8,0){\includegraphics[width=8cm]{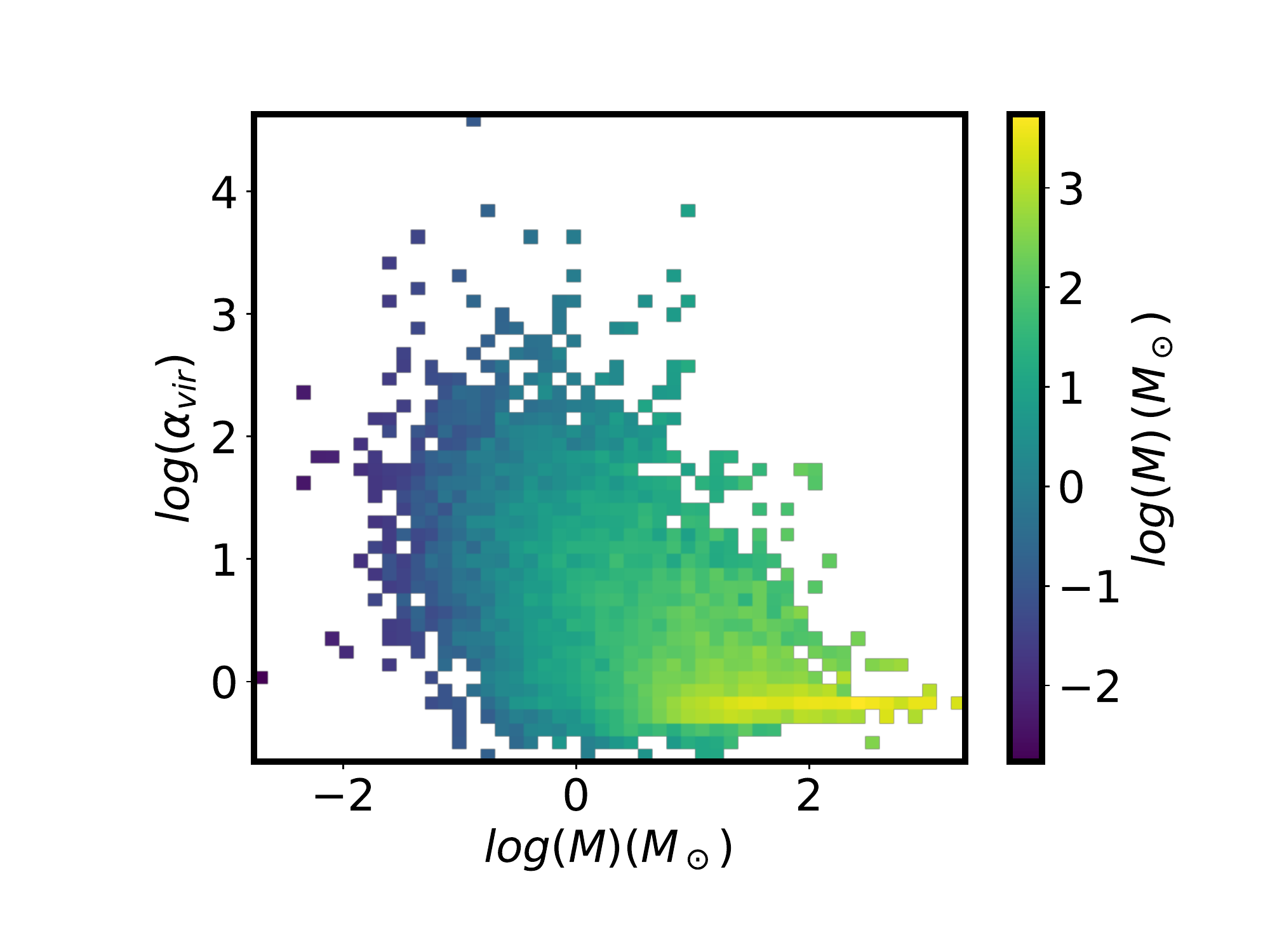}}  
\end{picture}
\caption{Physical properties of all structures identified in the simulation Z18 at time 10.04 Myr. 
Top-left shows the mass vs radius, top-right the mean density vs mass. 
Bottom-left shows  the  $\alpha_{vir}$ parameter (as stated by Eq.~(\ref{alpha_def})) 
for all structures while 
bottom-right one shows $\alpha_{vir}$ only for the structures with a mean density 
larger than 10$^5$ cm$^{-3}$. This latter confirms that most structures with 
a mean density larger than 10$^5$ cm$^{-3}$ are collapsed objects since $\alpha_{vir} \simeq 1$
(with little dispersion).  }
\label{struct_selec}
\end{figure*}

\setlength{\unitlength}{1cm}
\begin{figure*} 
\begin{picture} (0,17)
\put(0,11){\includegraphics[width=8cm]{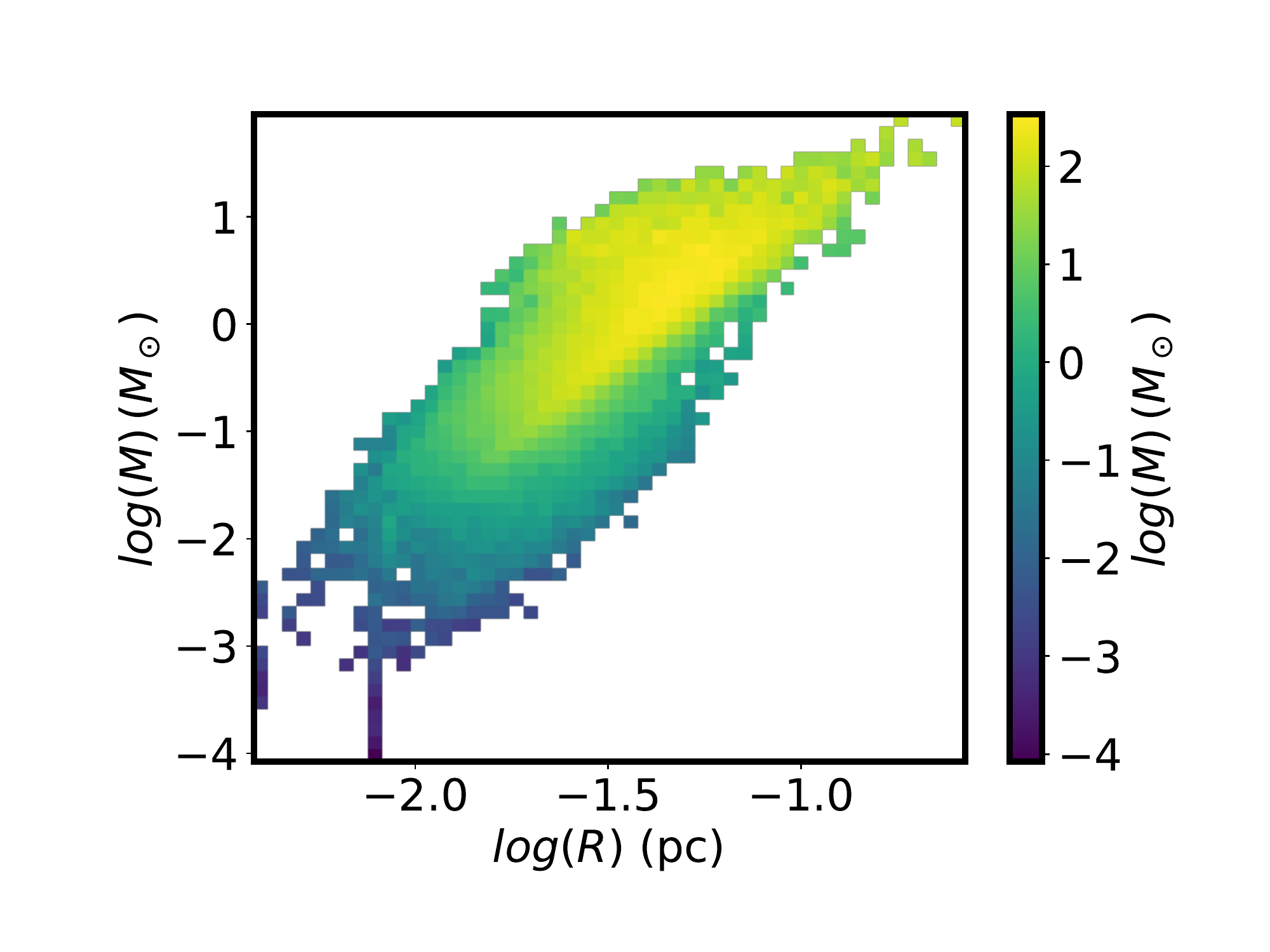}}  
\put(8,11){\includegraphics[width=8cm]{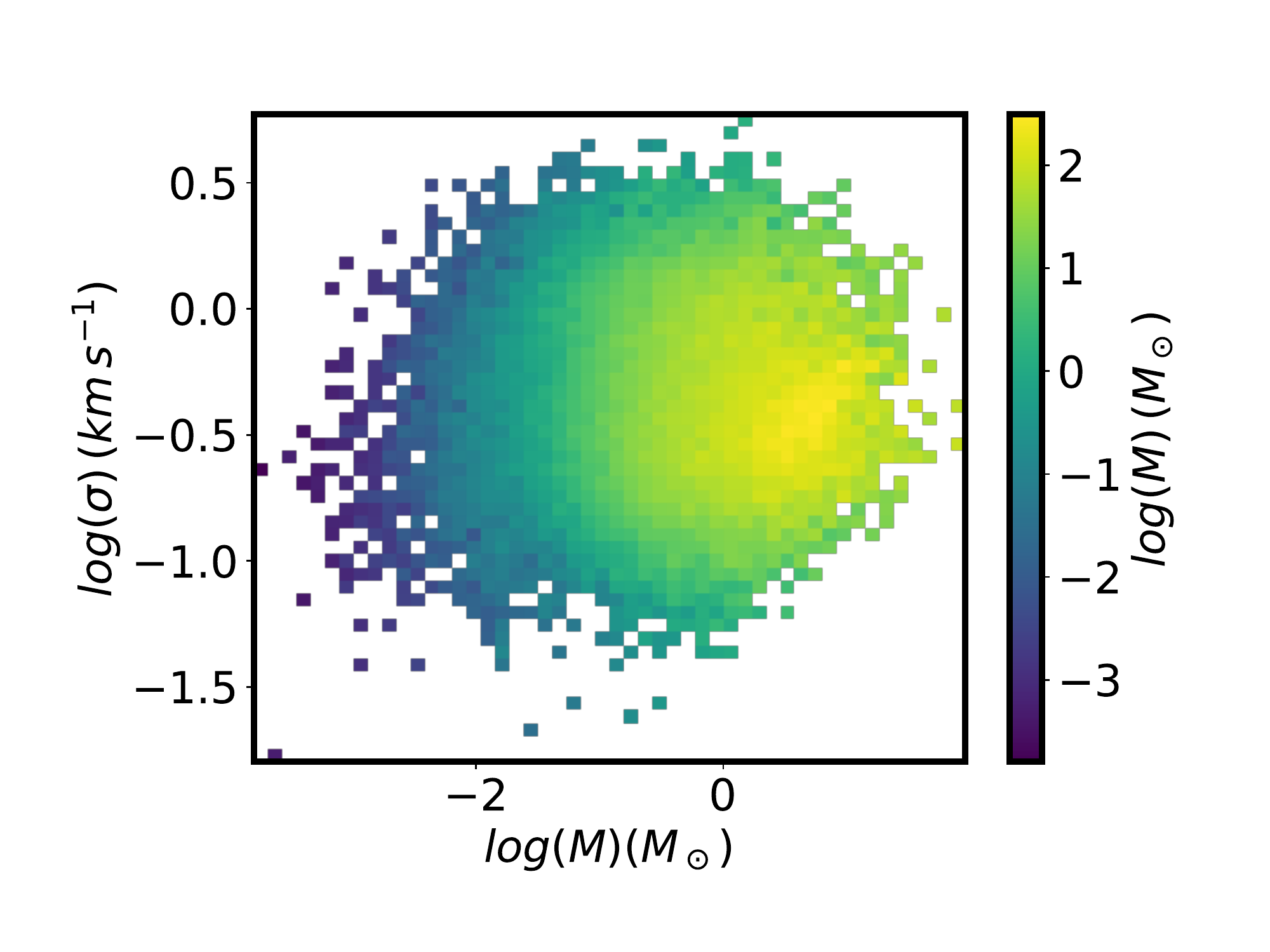}}  
\put(8,5.5){\includegraphics[width=8cm]{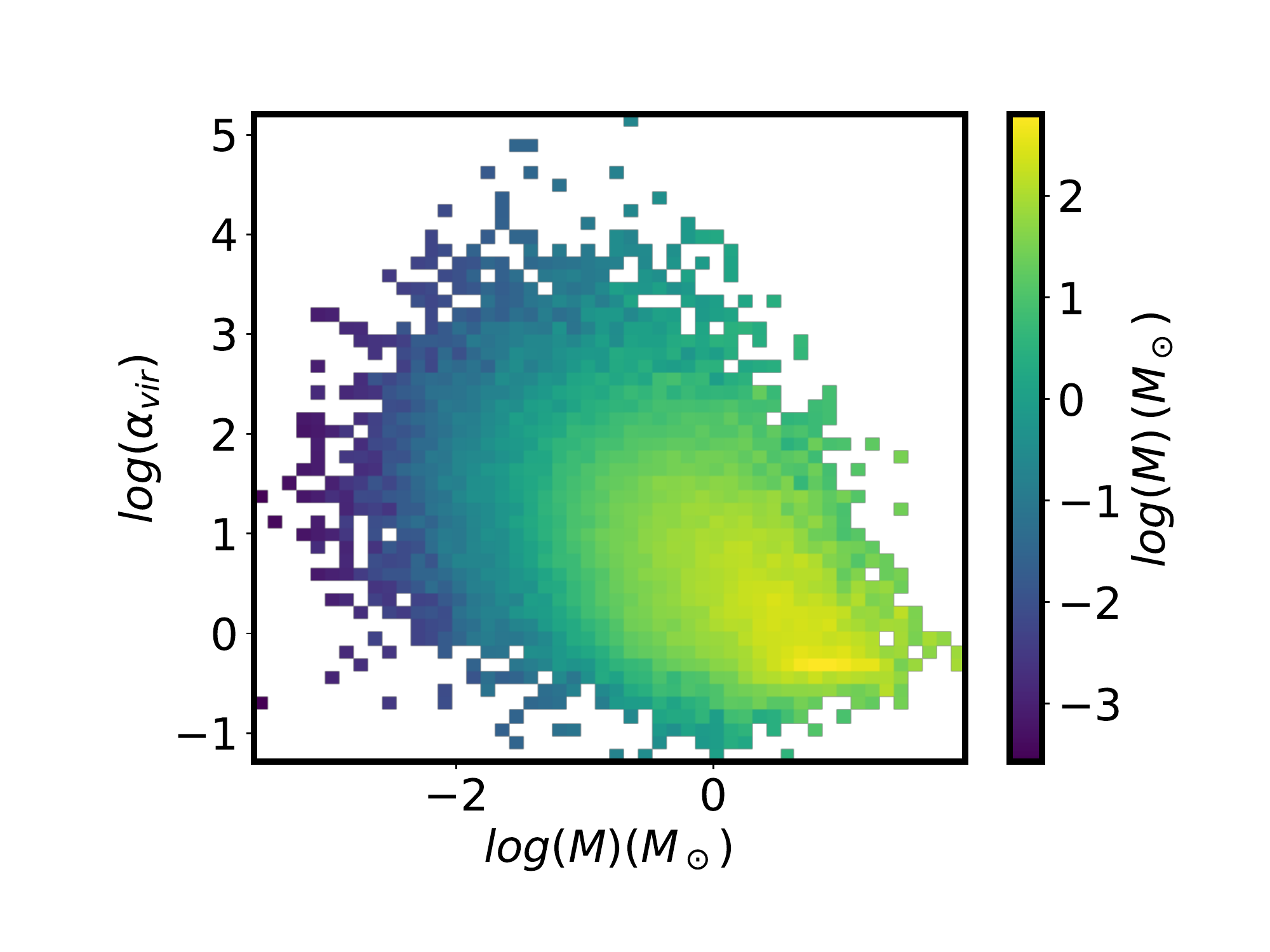}}  
\put(0,5.5){\includegraphics[width=8cm]{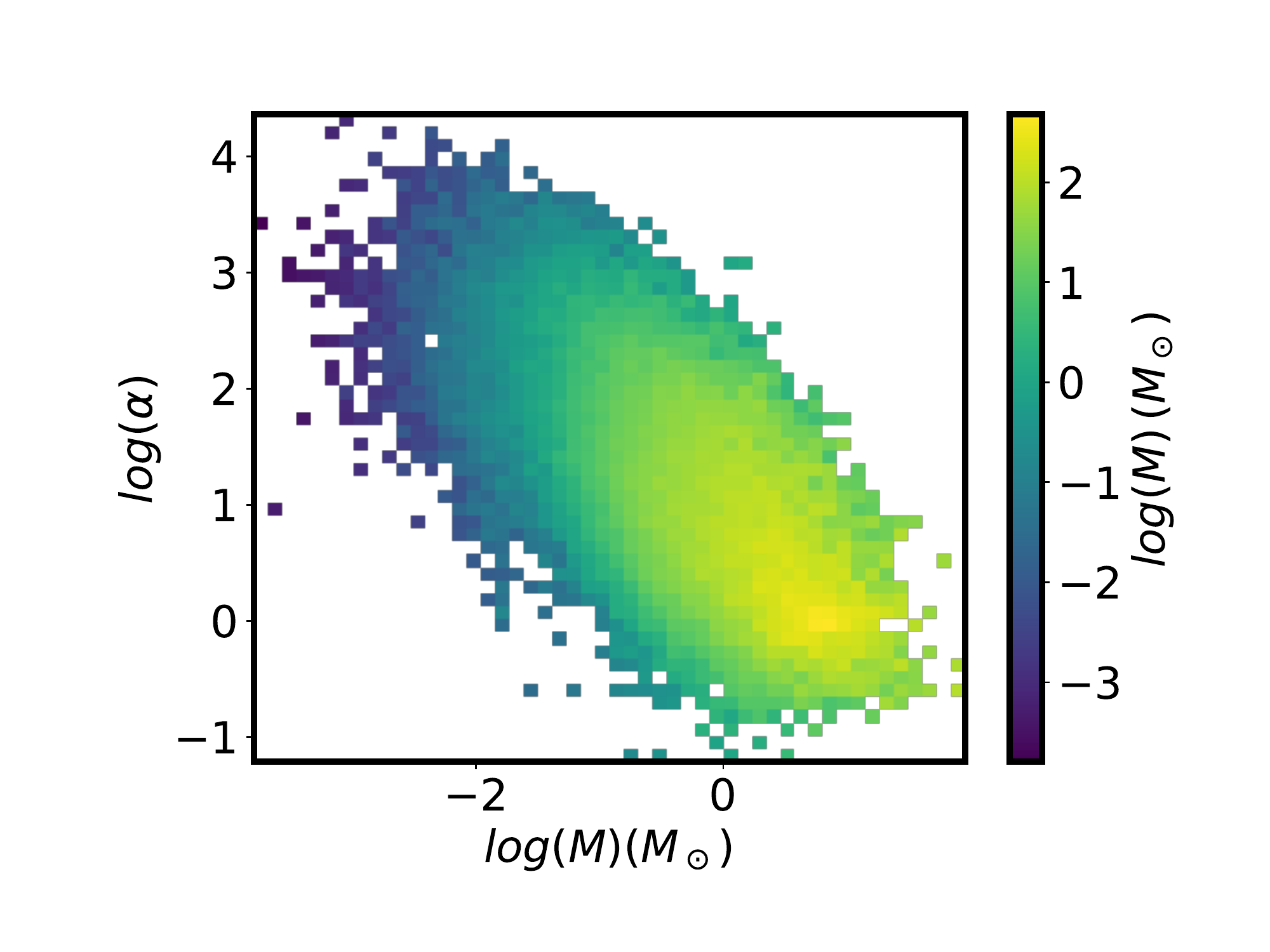}}  
\put(8,0){\includegraphics[width=8cm]{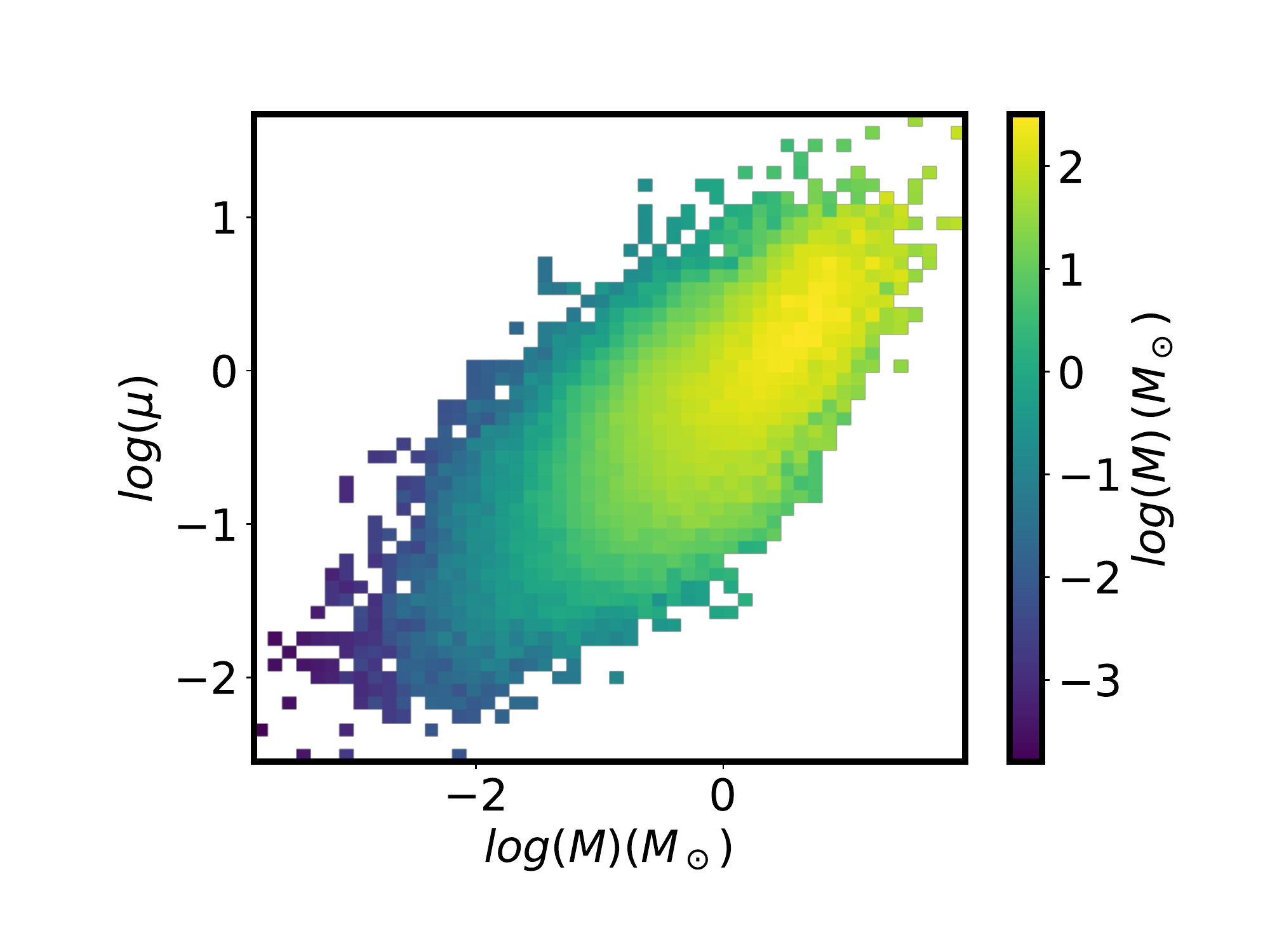}}  
\put(0,0){\includegraphics[width=8cm]{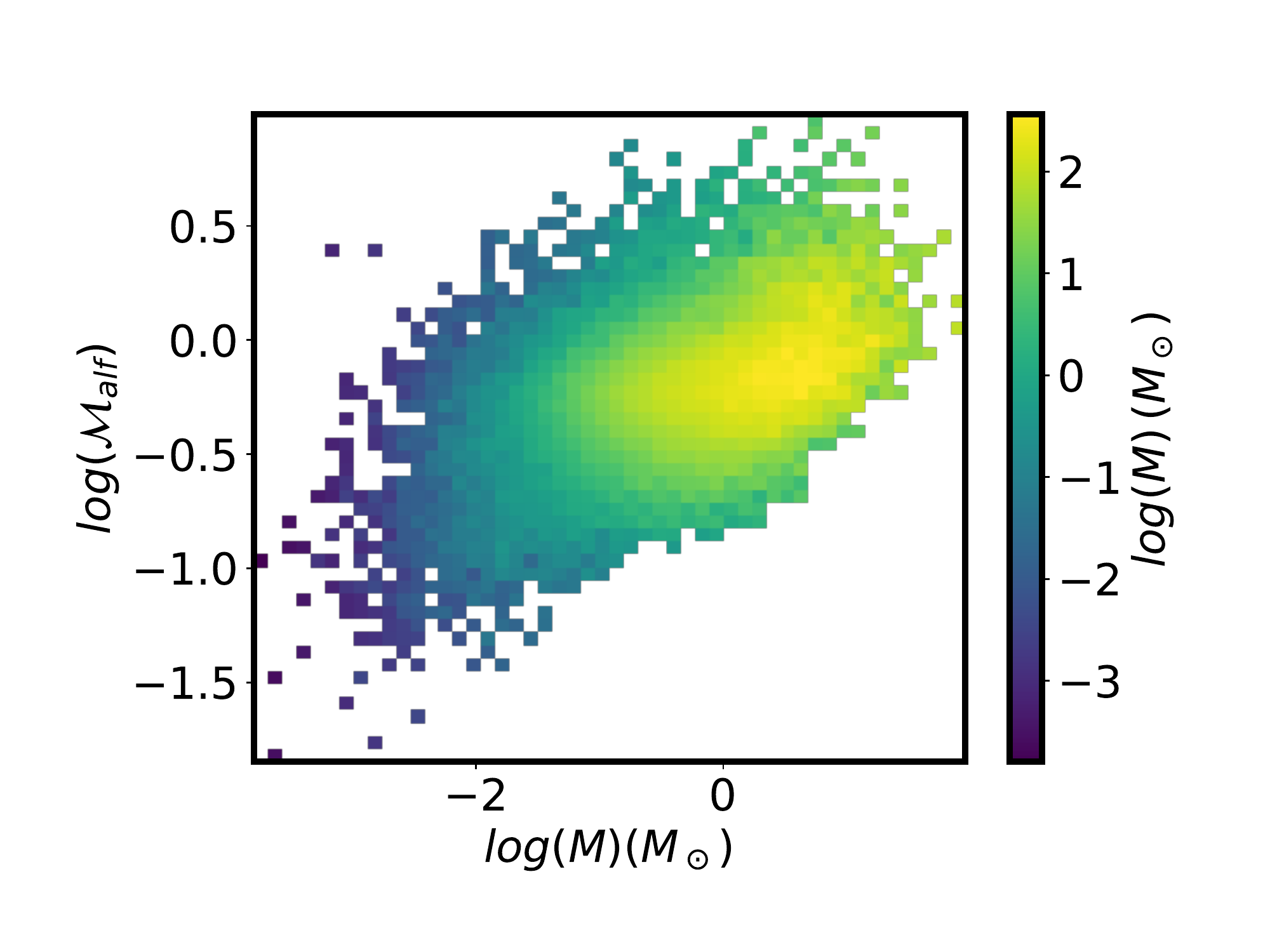}}  
\end{picture}
\caption{Physical properties of structures with mean density smaller than 10$^5$ cm$^{-3}$
in simulation Z18 at time 10.04 Myr
(as seen with Fig.~\ref{struct_selec} these structures are not dominated by collapsed objects). 
Left-top shows the mass-size relation (compare with top-left panel of Fig.~\ref{struct_selec} which 
shows the same quantity for all structures). Right-top shows the velocity dispersion 
(see Eq.~(\ref{def_stat1})) as a function of mass. Typical values are on the order of 
$\sigma \simeq 0.3$ km s$^{-1}$. Second row shows $\alpha$ and $\alpha_{vir}$. Most 
cores have values on the order of, or smaller than, a few.
The third row displays the Alfv\'enic Mach number and the $\mu$ parameter. 
 }
\label{struct_prop}
\end{figure*}

\setlength{\unitlength}{1cm}
\begin{figure*} 
\begin{picture} (0,18.5)
\put(0,12){\includegraphics[width=8cm]{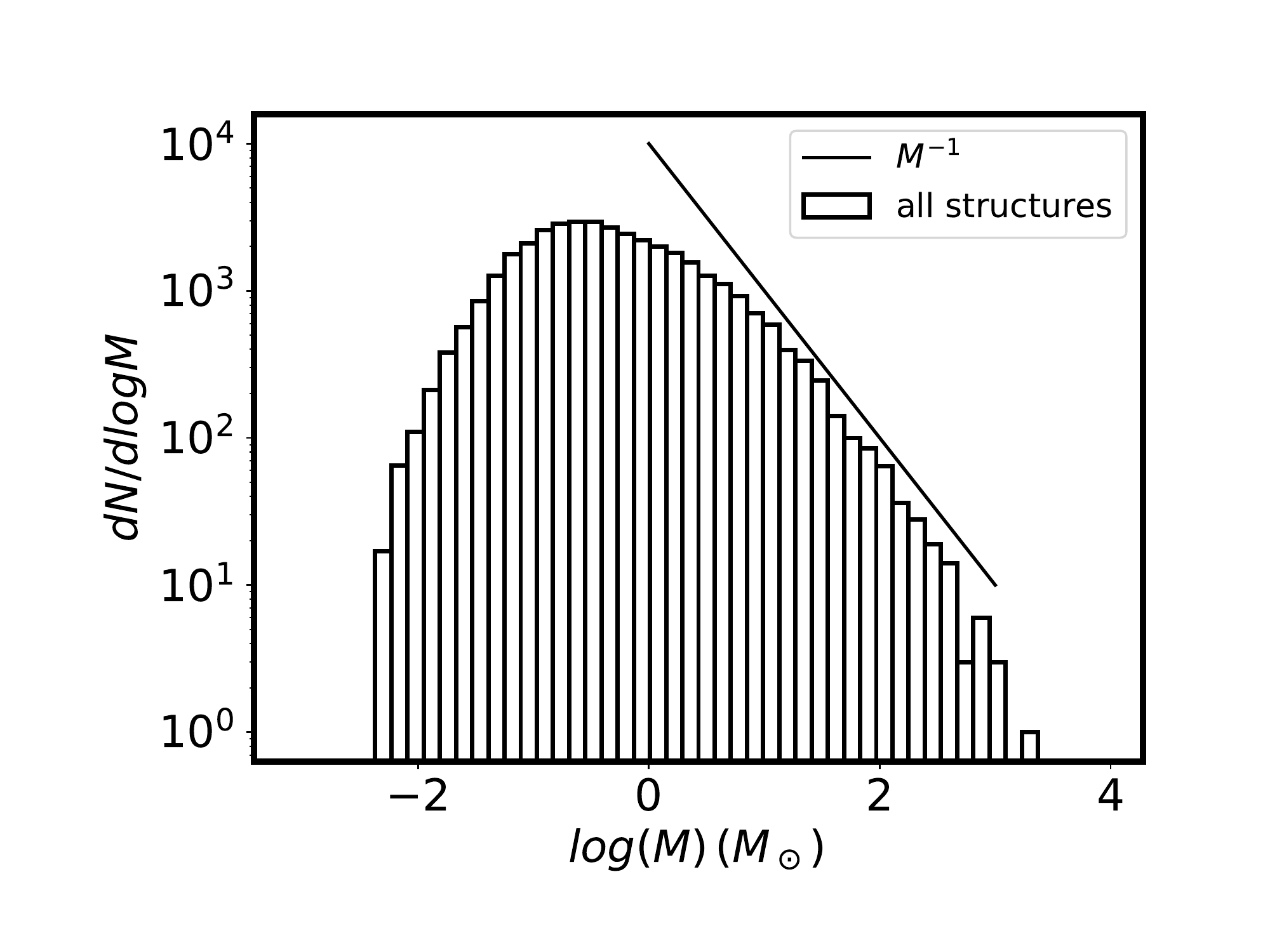}}  
\put(0,6){\includegraphics[width=8cm]{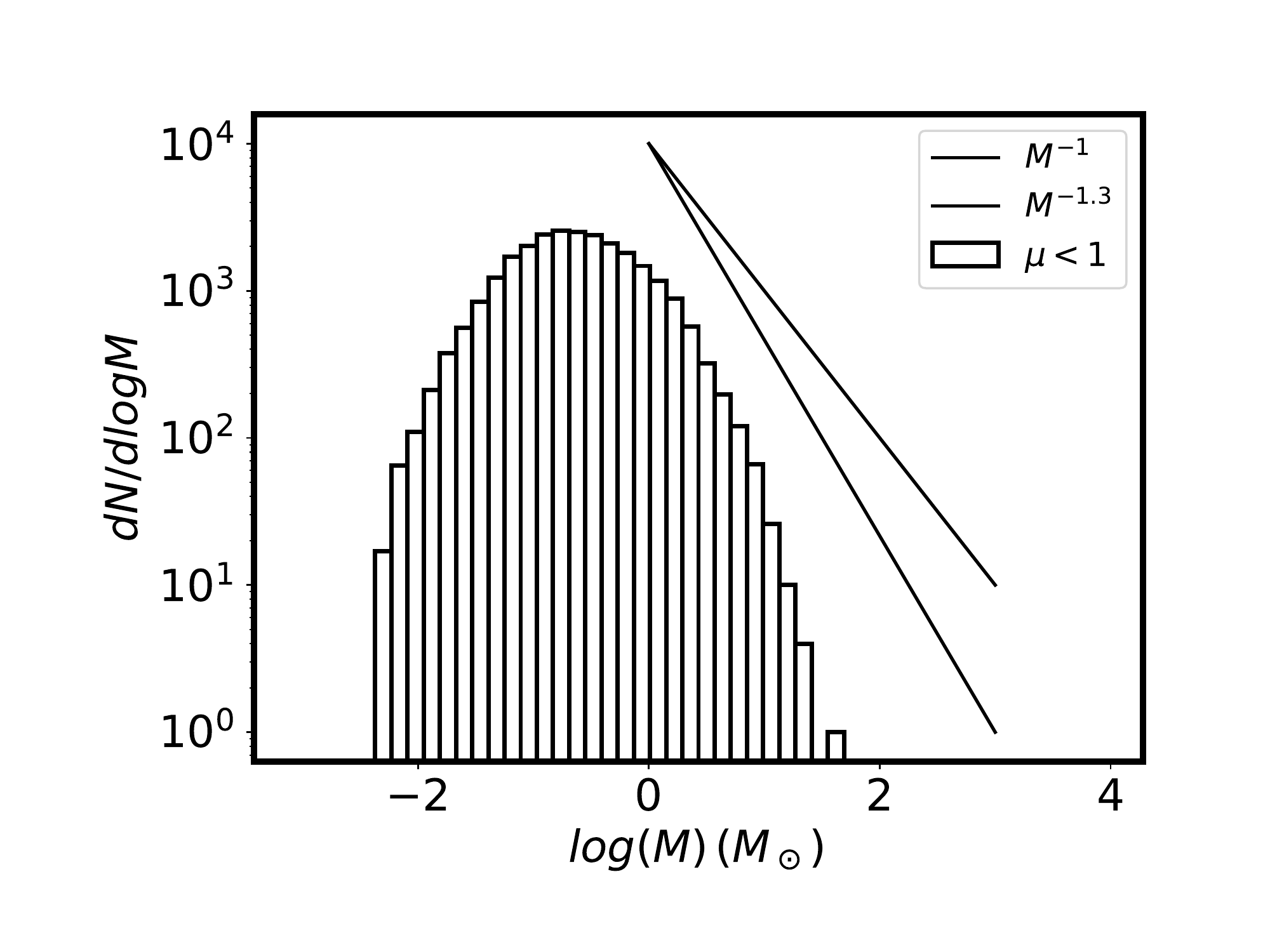}}  
\put(9,6){\includegraphics[width=8cm]{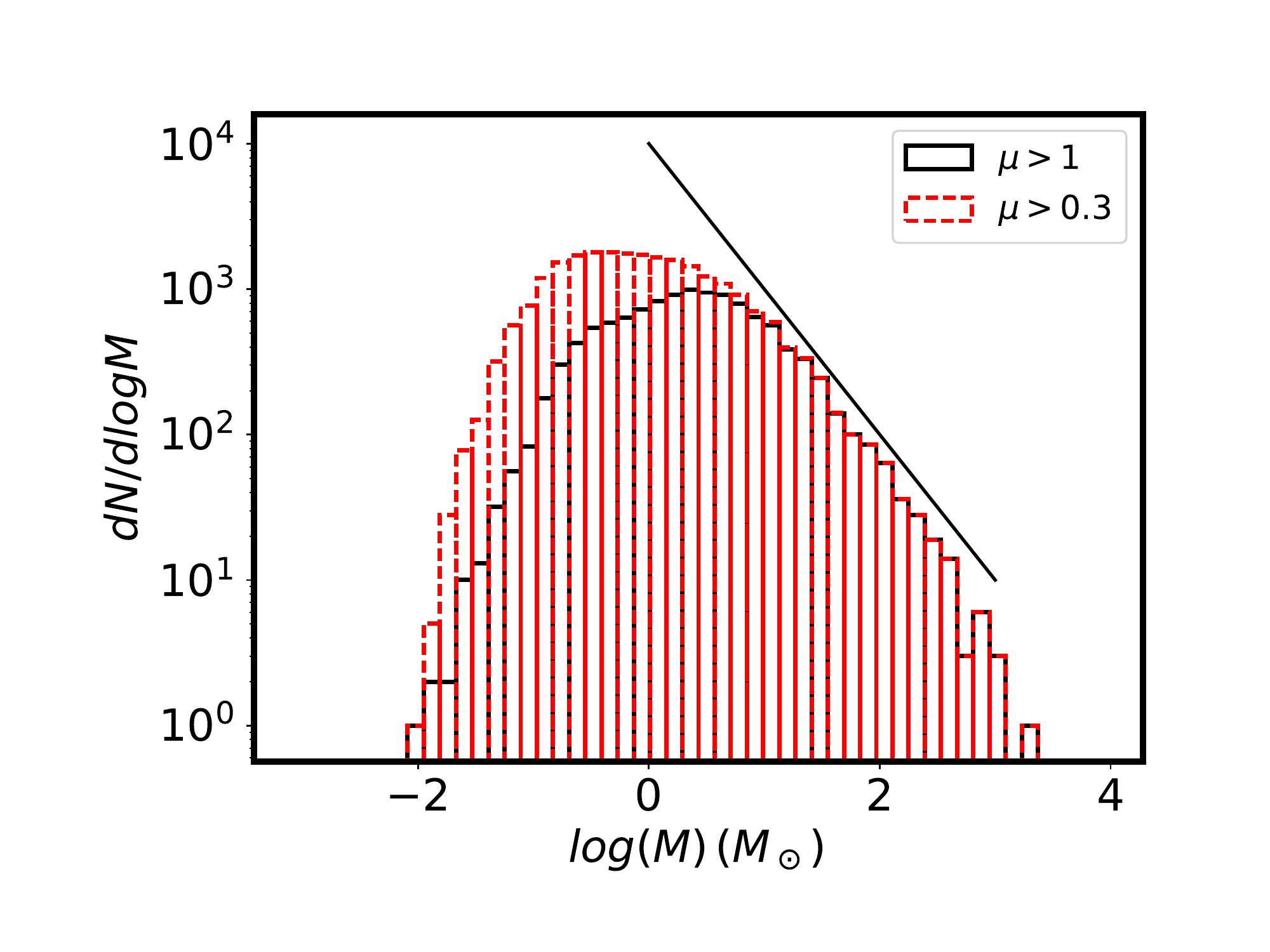}}  
\put(0,0){\includegraphics[width=8cm]{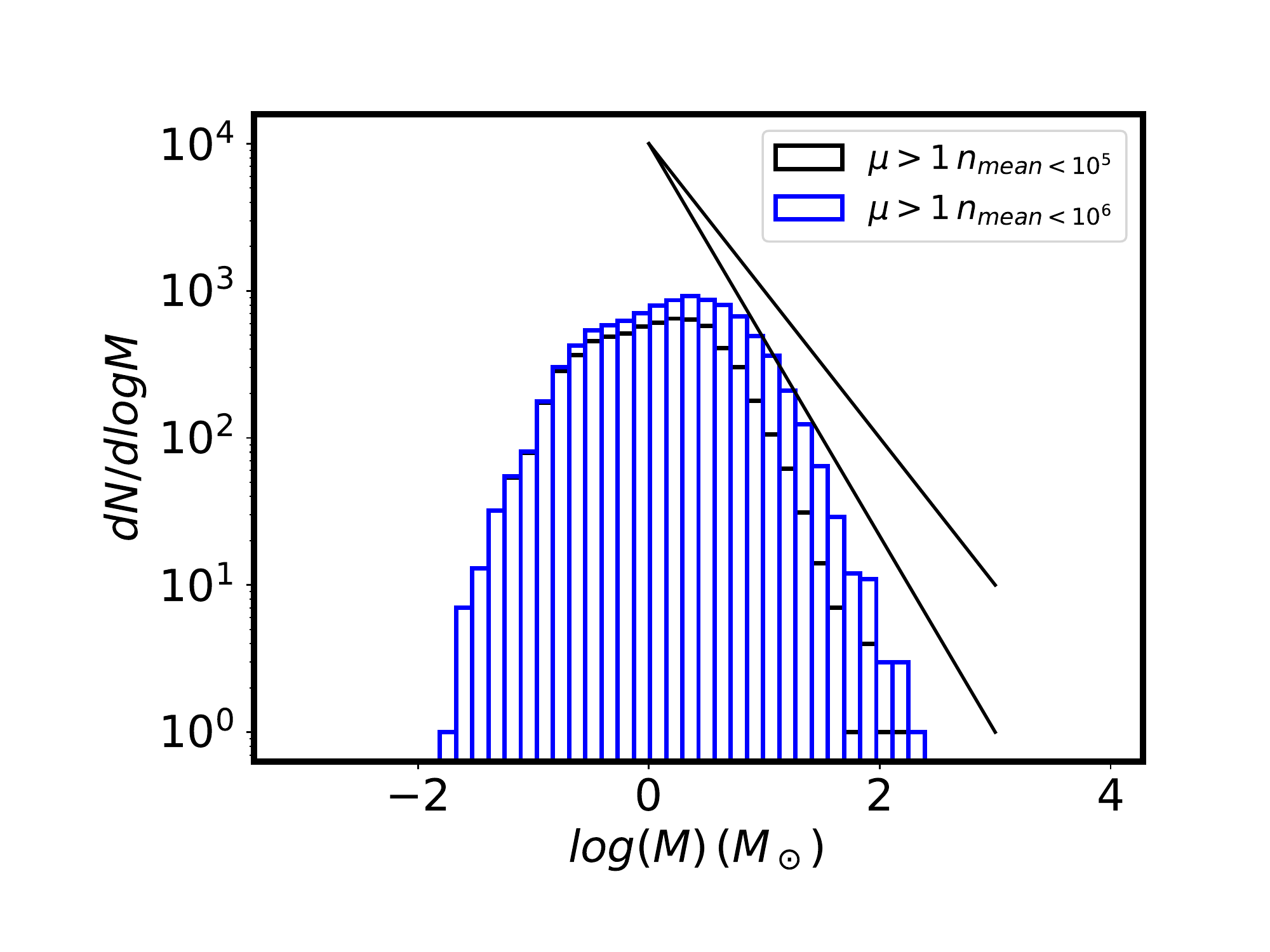}}  
\put(9,0){\includegraphics[width=8cm]{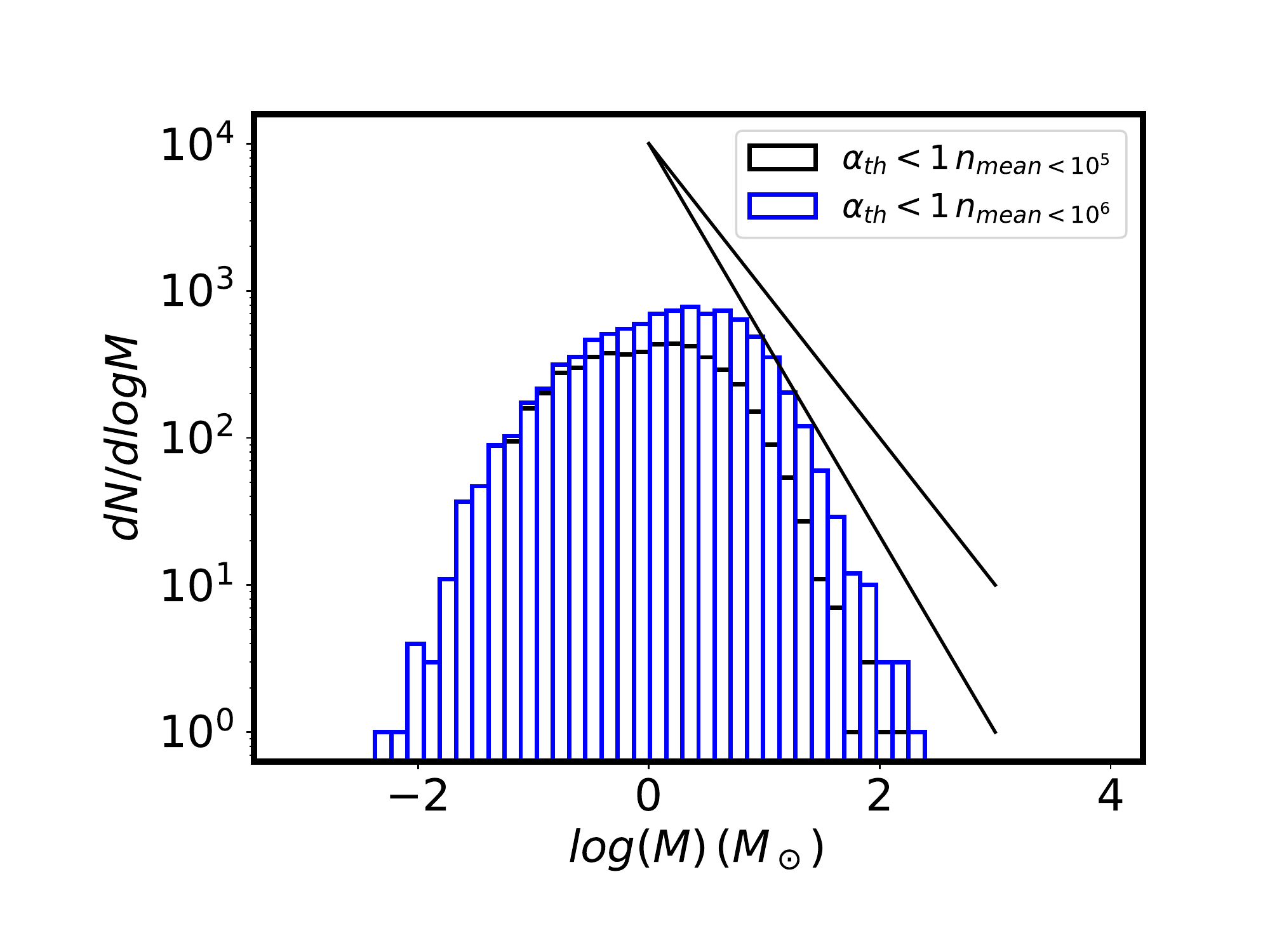}}  
\end{picture}
\caption{Mass spectra of the extracted {\it cores} at time 10.04 Myr in simulation Z18.
Top panel: no selection applied. Second panel: only subcritical {\it cores}, that is 
to say those with $\mu < 1$ are displayed. Third panel supercritical 
{\it cores} ($\mu > 1$, solid line) and {\it cores} with $\mu > 0.3$ (red dashed line). 
Bottom panel supercritical cores with central densities smaller than 
$10^5$ cm$^{-3}$ and $10^6$ cm$^{-3}$ (respectively black and blue solid curves).
 }
\label{mass_spectra}
\end{figure*}

\subsection{Strategy for zooming simulations}

The primary goal of the present study is to link the intermediate scales of
galaxies, that is to say the scales on the order of 100 pc-1kpc,  with the ones of the dense cores, 
thought to be the  mass reservoir  of stars. Dense cores have typical size on the order 
of, possibly slightly below, 0.1 pc \citep{ward2007,konyves2015}. To properly describe this scale, it is necessary to 
use  at the very least 10 cells across the cores  and thus to reach a spatial resolution of at least 10$^{-2}$ pc, 
which would give roughly 600 grid cells in a sphere of radius equal to 0.05 pc. 
On the other hand a reasonable description of the intermediate galactic scales require 
to describe typically of computational box of 1 kpc with at least 256 cells \citep{kim+2013,hennebelle2014,gatto2015,iffrig2017}, 
leading to a spatial resolution on the order of 4 pc.  
Clearly to make the connection between the few pc scales and the $10^{-2}$ pc ones, requires intense zooming. To handle this issue, 
we proceed as follows. 

First  we perform a  supernovae regulated ISM  simulations as described in  \citet{hennebelle2014} and \citet{iffrig2017}. 
For that purpose we use a grid resolution of 512$^3$. We run it for about 32 Myr, which is typically what is required to
obtain a multi-phase ISM self-consistently generated by supernova explosions. By this time about 1000 supernova explosions 
have occurred. Note that unlike what is done 
in  \citet{hennebelle2014} and \citet{iffrig2017} we do not use sink particles because at this resolution of a few pc, 
they represent large ensemble of stars (with masses on the order of 10$^{4-5}$ M$_\odot$) 
rather than single stars and they would affect the calculation onto the refined grids.
Therefore in these simulations, we prescribe a supernova rate. 
 Since the supernova rate in the Milky Way is about 1/50 yr$^{-1}$ and since most supernovae explode 
within the central 8 kpc, we take a supernova rate  of 1/50 yr$^{-1}$/$(\pi \times  8^2) \simeq 10^{-4}$ yr, 
which is therefore  roughly equivalent  to the Milky Way one for a region of 1 kpc$^2$. The supernovae are placed randomly in a sphere 
of 10 pc around the densest cell in the simulation. This scheme is therefore very close to the scheme ``C'' described 
in  \citet{hennebelle2014} except that the supernova rate is not temporally correlated with the star formation rate. 
Let us stress that with this approach, supernovae explosions start more rapidly than when sink particles are used,
 since there is no need to get collapse to generate them. This also implies that statistical equilibrium is reached faster.

At time $t=32$ Myr, we start zooming in a particular region. We increase the resolution on a square of size half the total box
length and we perform a few tens of timesteps (of coarse levels) in order to let the small scales relaxing and adapting to the new resolution. 
We repeat this procedure four times, that is to say increasing the resolution by a factor two on a region two times smaller
and performing a few tens of timesteps before increasing the resolution again. Note that   the size of the zooming region is enforced to 
be  at least 100 pc since the goal is to get enough statistics. To optimize computing resources, we have derefined the 
cells outside the first region of zooming bringing them to level 7 instead of 9. 
By doing so, we avoid too steep resolution jumps and we insure uniform resolution on the 
regions of interest, which optimizes the treatment of turbulence. In terms of resolution this corresponds 
to a cell size of about 0.06 pc. 

Finally, we allow for  further refinements up to four more AMR levels (for the fiducial run, see below), 
based on the Jeans length criterion being 
described by at least 10 cells.  To avoid increasing significantly the refinement too rapidly, we first allow for two 
levels of refinement and run the simulation during about 1.5 Myr which correspond to a few freefall and crossing times 
for gas densities of $n \simeq 10^3$ cm$^{-3}$. 
Altogether the simulation run about 5.6 Myr between 
the end of the unigrid calculation and the beginning of the full resolution calculation. 
These numbers are similar to the ones quoted in \citet{seifried2017}.

This provides (for the fiducial run) a finest spatial  resolution of 0.0038 pc implying that the scale of 0.1 pc is solved by about 
25 cells. A core of diameter 0.05 pc  contains about 9000 cells. 
While such a type a resolution is not sufficient to describe the details within collapsing cores
\citep[e.g.][]{masson2016,h2016}, it is sufficient to identify the cores and infer their mass.
The resulting mesh is illustrated in Fig.~\ref{zoom_series}, that shows a series of zoom illustrating the 
high resolution dynamics. Top-left panel shows the  maximum AMR levels along the z-axis.
Note the first four levels of uniform refinement and the four further ones based on Jeans criterion 
and therefore centered around column density peaks.

Let us stress that in this work, we do not use sink particles, even when full resolution is achieved,
 as the spatial resolution is still not sufficient to provide a 
description of individual stars and sinks on the order of few tens of solar masses would be obtained. 
Note also that once we start refining we stop introducing supernovae remnants because the combination of 
very high velocities (on the order of a 100 km s$^{-1}$) and the high spatial resolution leads prohibiting low 
timesteps.  In any case, since massive stars have a life time larger than 4 Myr, supernovae are not 
expected to have a very strong impact in dense star forming regions because they come too late. Moreover
other types of feedback such as ionizing radiation should in principle be considered \citep[e.g.][]{geen2017}.

\subsection{Runs performed}
The influence of several aspects of the procedure we used, needs to be investigated. 
On the other hand the runs are quite expensive (typically several millions of CPU hours) and only a 
few can be carried out. 

We believe that the most important parameters are the maximum resolution and the influence of 
the time at which the zooming is performed. To tackle these questions we have performed 
three runs as described in table~\ref{table_run}.

      \begin{table}
         \begin{center}
            \begin{tabular}{lccc}
               \hline\hline
               Name & run time (Myr) & $l_{max}$ &  Resolution (pc) \\
               \hline
               Z17 &   2.4  & 17    & 0.0072 \\
               Z18 &   4.3  & 18    & 0.0036 \\
               Z19 &   0.8  & 19    & 0.0018 \\
               \hline
            \end{tabular}
         \end{center}
         \caption{Summary of the runs performed. The three runs start from the same point. The run time 
         is the duration of the numerical simulation. $l_{max}$ is the maximum AMR level used in this simulation and 
the resolution, the physical scale of smallest computational cells.}
\label{table_run}
      \end{table}
The runs have been performed on 4000 cpu and have typically several hundreds of millions of computing cells (depending on resolution and time).
Altogether they have requested about 10 millions of cpu hours. 

By comparing the results of the three runs (Sect.~\ref{convergence}), we will be able to 
quantify the impact of the resolution, which is a key aspect. Simulations Z17 and particularly Z18, 
which is our fiducial run, have been performed for a few Myr. This corresponds to the freefall time for
 densities of about 100 cm$^{-3}$. Therefore for these 2 simulations, the most recent collapsed objects are made 
from gas that was diffuse enough by the time the zooming started. Thus by looking at the evolution of the 
structure properties, we can infer to what extent their properties are affected by the time and also the 
resolution of the simulation just before the zooming starts. Anticipating over the  results of 
section~\ref{time_evol}, we find that the statistics appear to be robust to time evolution, 
seemingly suggesting that the starting point at which zooming is performed is not too severe an issue. 

Note that because of computing power limitations, the Z19 simulation could not be run for longer time.
However since  the evolution of statistics with time remains limited in the Z18 run (see section~\ref{time_evol}), in principle
this corresponds to a sufficiently long time to get stationary statistics. 

\subsection{Missing physics}
There are numerous important processes, which are not included in this work. While 
we believe, it is important to proceed step by step to decipher their respective impacts, we 
briefly and  qualitatively recall their possible effect. 

First of all, we assume ideal MHD, that is to say we do not model the ion-neutral friction 
which probably has an impact on the core formation  \citep{vanloo2008,kudoh2008,kudoh2011,chen2014} and
the turbulence \citep{li2008,tilley2011,burkhart2015,ntormousi2016}. This implies that at the core scales,
the magnetic field structure could possibly be smoother and the magnetic intensity lower that what the simulation predicts.

Second of all once refinement starts, we do not include any stellar feedback 
that would $i)$ limit star formation by disrupting molecular clouds 
through ionizing radiation \citep[e.g.][]{walch2012,dale2013,dale2014,geen2015,geen2016,geen2017}, 
$ii)$ modify the core distribution as it has been reported for example for the 
jets \citep[e.g.][]{wang2010,federrath2015}. These effects may modify the statistics by 
generating a second generation of cores whose formation has been triggered, or at least influenced
by the feedback of the first generation.

\subsection{Qualitative description}

Figure~\ref{zoom_series} shows a series of zoom from the kpc box (right-top panel)
to a few 0.1 pc (right-bottom panel). Right-top and left-second row  panel show the stratification 
along the z-axis. The typical thickness is about 50-100 pc (depending of the gas 
density, see \citet{iffrig2017} for a detailed discussion).  Visually the 
aspect of the gas looks broadly similar from scale of 50 pc (right-second row panel) to 
10 pc (left-third row panel). We see that the medium is highly structured with clumps at all scales
and very prominent filaments also at all scales. This is less the case at scale of about 
3 pc (right-third row panel) and even less for bottom panels. This behaviour is 
possibly a consequence of gravity becoming more and more important within the selected regions 
(concentric cubes around $x=133.9$ pc and $y=502.8$ pc) while turbulent energy tends on the contrary 
to be smaller and smaller (because of its scale dependence).
 It  may also indicate that 
the small scales are not completely described since Jeans length based refinement instead of uni-grid is being used
for the four last levels, as discussed above. 

The two bottom panels show that the dense gas is very fragmented in relatively well 
defined cores. Some of them, however, show signs of interactions or complex 
morphologies as seen in right-bottom panel. At this point, it may be difficult to decide
whether this should be described as a single core with a complex inner structure or 
as two interacting cores. 
In the rest of the paper we describe how these cores are being defined  and 
we study their statistics.

\section{Structure extraction and properties}

\subsection{HOP algorithm}
The main goal of the present paper is to study the prestellar cores  in the context of a self-consistently
generated ISM and we must proceed to their extraction. 
For this purpose, we use the group finding algorithm, HOP, which has been 
widely used in the cosmological context to detect dark matter haloes \citep{eisenstein1998}. 
This algorithm is also used in the ISM context by \citet{bleuler2014} to identify the possible 
location of new sink particles.
HOP associates to each particle
its densest neighbour, repeating the procedure this defines a path which ends when the particle is its own densest neighbour.
The ensemble of particles, which end at the same local density maximum are called a group. There are few users parameters 
that have been found to have little influence on the final result with the notable exception of the density
threshold  above which particles are considered \citep{eisenstein1998}. Once the groups are obtained the algorithm
also offers the possibility to merge the groups, something that we do not use in the present study.

To use HOP we proceed as follows. First, we select in the simulation all the cells that have a density above 
3000 cm$^{-3}$, located inside the maximally refined regions (corresponding to the green square visible in 
top-left panel of Fig.~\ref{zoom_series}). These spatial coordinates and the density of these cells
 are then provided to the HOP algorithm, which groups them following the procedure described above. 
Note that as discussed below, most structures found this way are not self-gravitating and should not be
classified as cores, a point to which we come back below where several criteria are being studied. 
The word {\it cores} will refer to {\it structures} (i.e. groups of cells identified by HOP) which satisfy 
a specific criterion (typically based on virial analysis). \\

Note that at this stage, we do not attempt to define and extract the cores as the observers
do. The reason is that this is in itself a challenging process, which requires several steps
including a modelisation of the observations themselves as well as the usage of specific software
\citep{menshchikov2012}. This goes beyond the scope of the present paper, which focuses on the 
method and the physical analysis of the structures formed.

\subsection{Computed quantities}
\label{comp_quant}
Once we get the groups of cells, we calculate the mass $M$,  
 the velocity dispersion, $\sigma$, the cloud radius, $R$, 
the virial $\alpha$ parameter 
   and the mass-to-flux over critical mass-to-flux ratio \citep{mouschovias1976}, $\mu$. 
For  some of these parameters, there are several possible choices. 
The spatial coordinates used in the equations below are with respect to the center of mass
of each individual structure.

The internal velocity dispersion is defined as
   \begin{eqnarray}
          \label{def_stat1}
     {\bf v_0} = {\sum {\bf v} \rho dx^3 \over \sum \rho dx^3 }, \\
     \nonumber
     \sigma ^2 = {1 \over 3} {\sum ({\bf v} - {\bf v_0})^2 \rho dx^3 \over \sum \rho dx^3 } , \\
   \end{eqnarray}


To define the radius, we first compute the inertia matrix
   \begin{eqnarray}
     I_{ij} = \sum x_i x_j \rho dx^3,
   \end{eqnarray}
that we diagonalise giving three eigenvalues $\lambda_i$. We then define
   \begin{eqnarray}
     R = \left( {\lambda_1 \lambda_2 \lambda_3 \over M^3 } \right)^{1/6}
   \end{eqnarray}

To characterize the dynamical state of the structures, we
 compute several values of the virial parameter, $\alpha$ as stated by Eqs.~(\ref{alpha_def}). First, we compute 
the standard observational definition that we will refer to as $\alpha$. Then we compute 
the exact ratio between the kinetic energy and the gravitational energy $\alpha _{vir}$. Finally, 
we also compute the ratio between the thermal and gravitational energy,  $\alpha _{th}$.
   \begin{eqnarray}
     \nonumber
     \alpha = {5 \sigma^2 R \over G M}, \\
     \label{alpha_def}
     \alpha _{vir}= { 2 E_{kin} \over E _{grav}} ={  \sum \rho  ({\bf v} - {\bf v_0})^2  dx^3  \over \sum  {\bf g_i . r_i} \rho dx^3  }, \\
     \nonumber
     \alpha _{th}= { 2 E_{th} \over E _{grav}} = { 3 \sum  P dx^3  \over \sum  {\bf g_i . r_i} \rho dx^3  },
   \end{eqnarray}
   where ${\bf g_i}$ is the gravitational field. 

   The mass-to-flux ratio is widely used to estimate the strength of the magnetic field with respect to gravity.  
   To compute $\Phi$, the magnetic flux, we first compute the cloud center of mass, then we compute the flux across the 
   three planes parallel to $xy$, $xz$ and $yz$ and passing through the center of mass. We then take the largest of these 
   three fluxes. 
   The exact definition of $\mu$, depends in principle on the object geometry and flux distribution, here 
   we use the definition of \citet{mouschovias1976}   
   \begin{eqnarray}
     \Phi = \sum B dx^2, \\
     \nonumber
     \mu = {M  \sqrt{G} \over 0.13 \Phi}.
     \label{mu_def}
   \end{eqnarray}

\setlength{\unitlength}{1cm}
\begin{figure} 
\begin{picture} (0,11.5)
\put(0,5.5){\includegraphics[width=8cm]{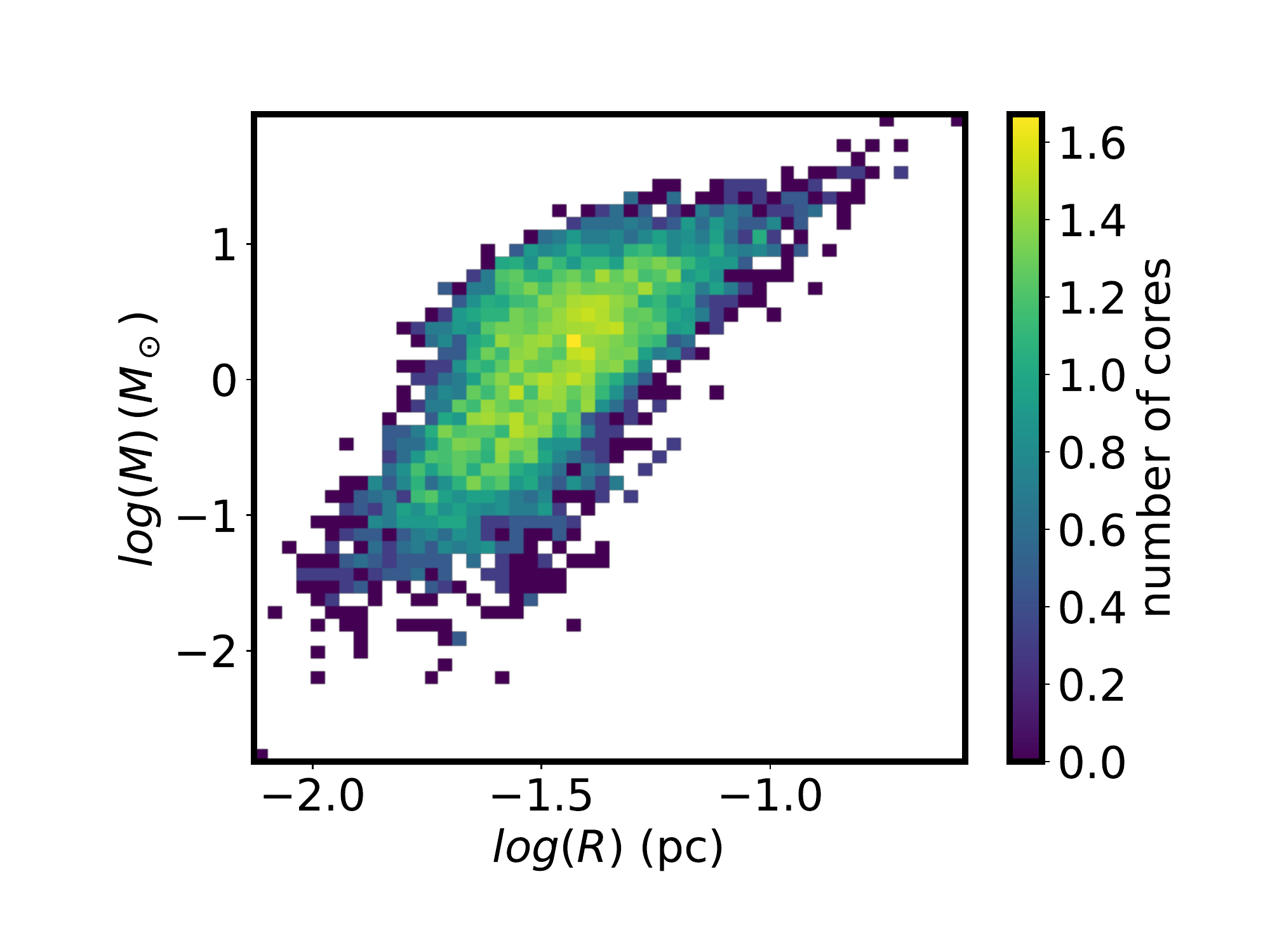}}  
\put(0,0){\includegraphics[width=8cm]{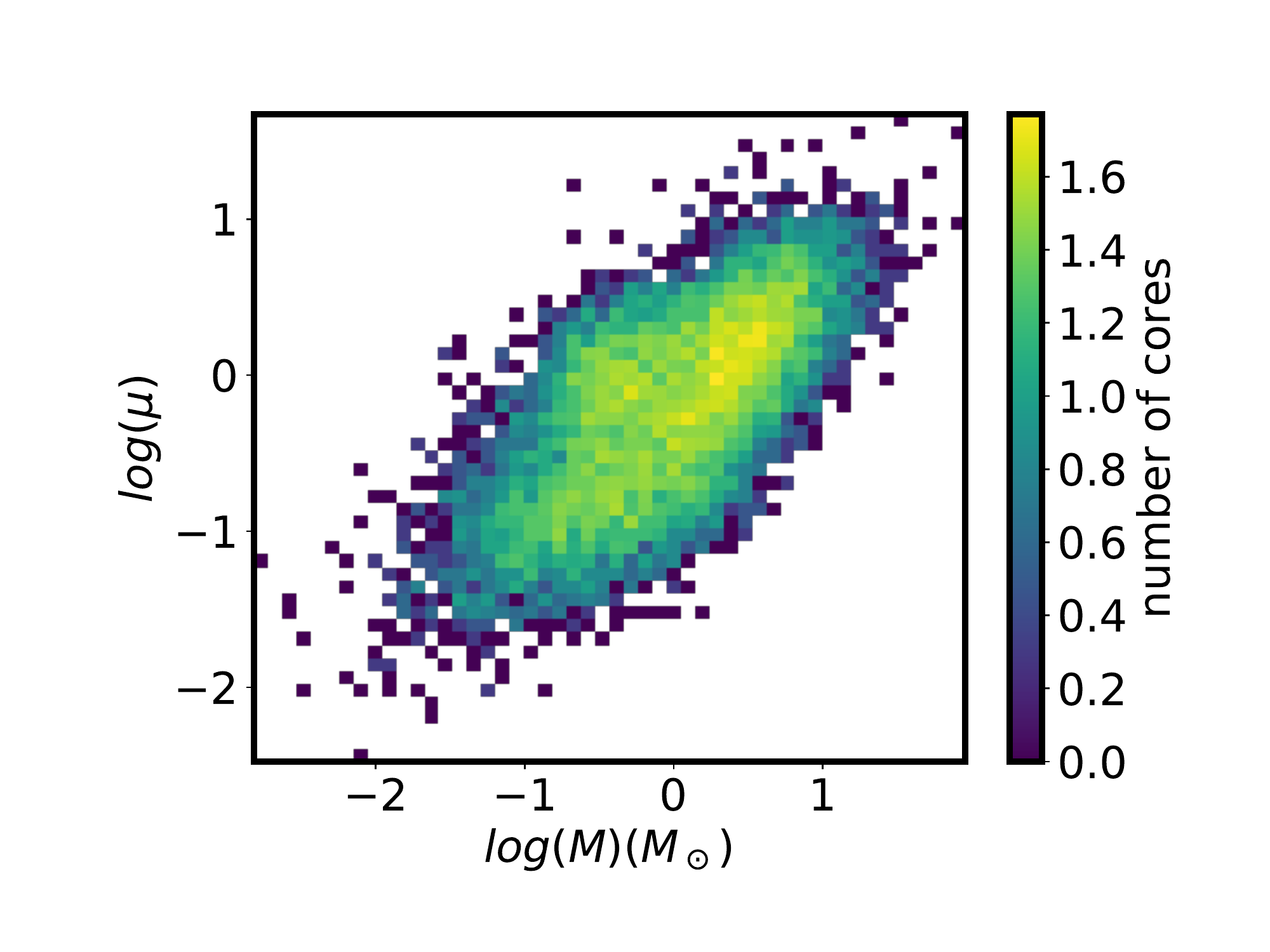}}  
\end{picture}
\caption{Upper panel: mass-radius relation of 
thermally supercritical cores.
Lower panel: magnetic mass-to-flux of thermally supercritical cores.
While massive cores are all magnetically supercritical, there is a 
significant number of intermediate and low mass cores, which are magnetically dominated. 
 }
\label{mu-th-vir}
\end{figure}

\setlength{\unitlength}{1cm}
\begin{figure} 
\begin{picture} (0,11.5)
\put(0,5.5){\includegraphics[width=8cm]{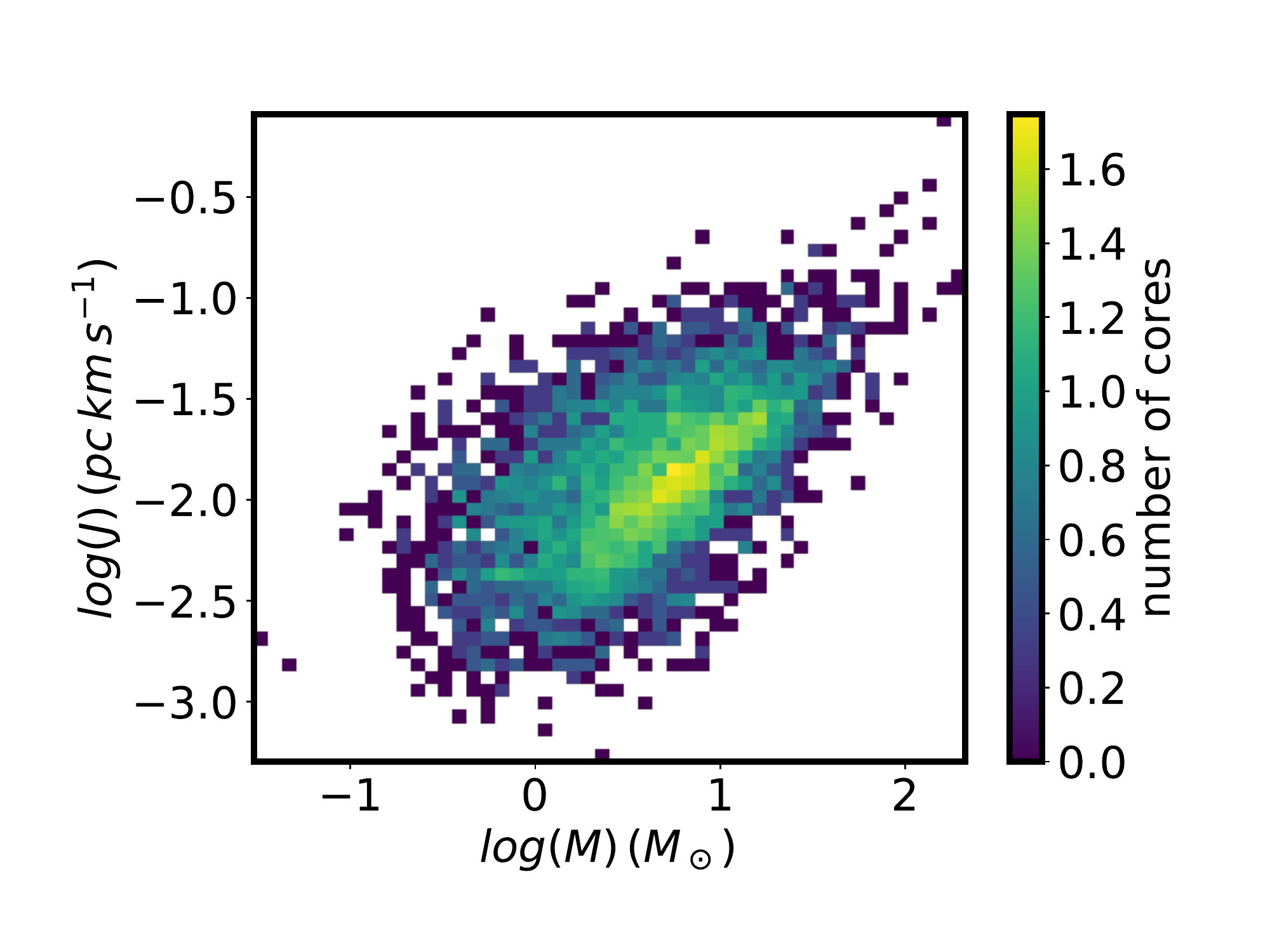}}  
\put(0,0){\includegraphics[width=8cm]{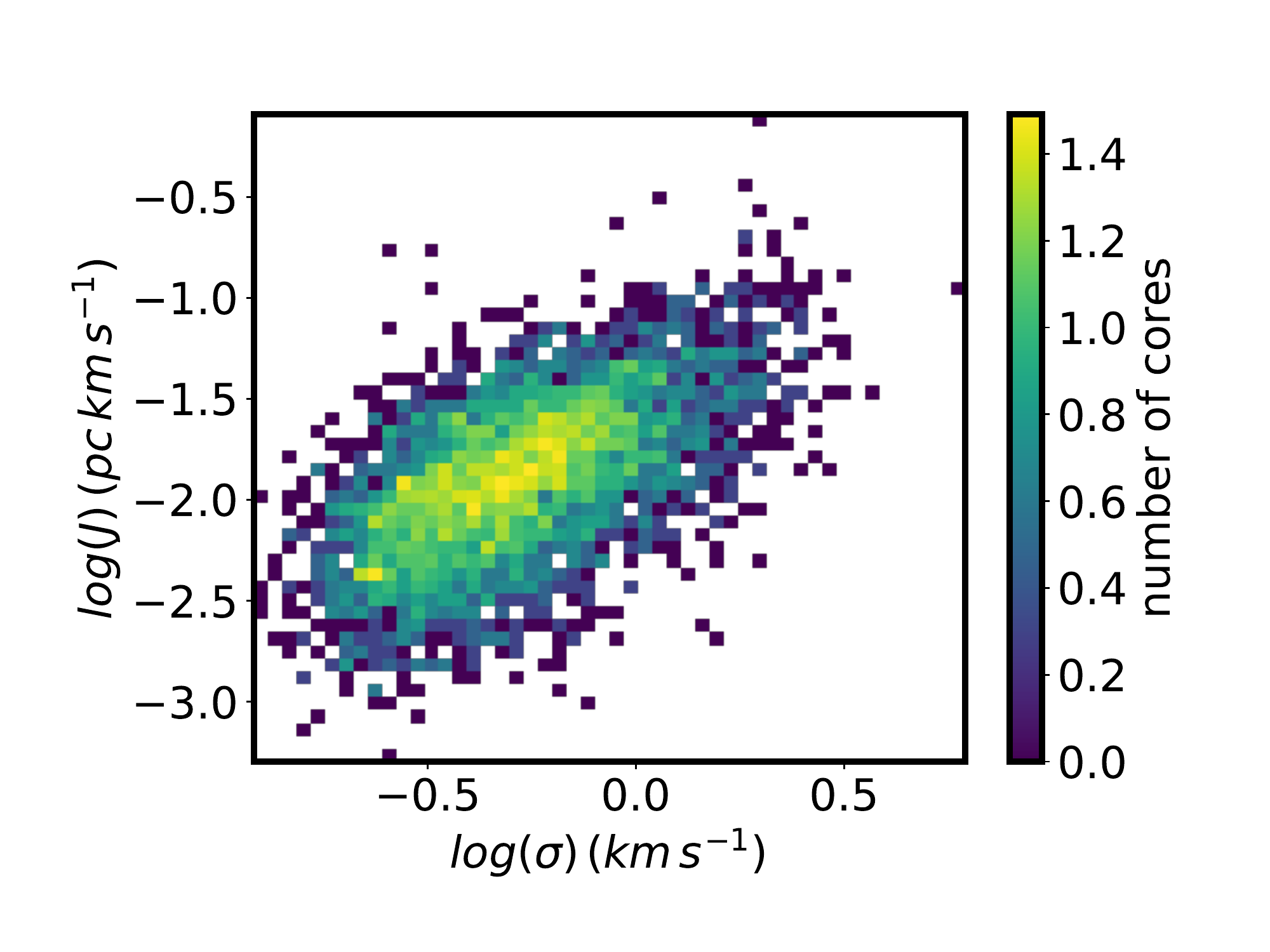}}  
\end{picture}
\caption{Distribution of angular momentum in magnetically supercritical cores
of mean density below $10^6$ cm$^{-3}$. Top panel displays it as a function 
of the mass while bottom panel gives it as a function of internal velocity dispersion.
 }
\label{ang_mom}
\end{figure}

\section{Statistical properties of cores}
We now turn to a description of the statistical properties of the extracted structures and cores.
In this section we present the results of run Z18, that is to say with a spatial 
resolution up to $3.6 \times 10^{-3}$ pc and  at  time 10.04 Myr.

\subsection{Mass, radius and density of structures: core selections}
Figure~\ref{struct_selec} shows a series of dimensional histograms displaying
various quantities  as described in section~\ref{comp_quant}. 
Top-left panel shows the mass as a function of the radius, $R$.  
While the radii span about one decade, from 0.01 to a few 0.1 pc,  the masses
vary over more than 4 decades reaching values below 0.01 $M_\odot$ and 
above 100 $M_\odot$. At first sight this could suggest that the 
radius  weakly varies with the mass. However, this is not exactly the case. 
 From the mass-radius distribution, the structures can be divided in two main populations.
First, a significant fraction of objects lies around a 
 line starting at $M \simeq 0.1 \, M_\odot$ , $R \simeq 0.02 $ pc 
and ending at  $M \simeq 10 \, M_\odot$ , $R \simeq 0.1 $ pc. 
This population of structures roughly follow $M \propto R^3$.
We call it region I. The second population is located around 
$M \simeq 10 \, M_\odot$, $R \simeq 0.02$ pc that we call region II. 

This second population corresponds therefore to objects much denser than the ones of 
the first population. This can be more clearly seen on the top-right panel that 
displays the mean density distribution. 
This latter is simply defined as the ratio of the mass structure over its total volume.
The structures of regions I have
  densities of about $10^{4-5}$ cm$^{-3}$ and  masses 0.1-10 $M_\odot$. 
The structures corresponding to regions II are at much higher density. 
This latter  
is nearly proportional to their mass. 

We believe that these  two types of structures should be distinguished. 
The first one represents structures which have not yet strongly collapsed such 
as pre and protostellar cores. The second type 
corresponds to objects which have collapsed and therefore,  since as explained 
above, we are not using sink particles, their mass has piled up on a few 
computing cells explaining why the density increases with their mass. 
These objects  therefore represent Young stellar objects (YSO). Note 
however that since merging is occurring, their distribution evolves 
with time and bigger objects are gradually built up. This indicates that 
as we are studying the statistical properties of cores, it is 
necessary to separate the two populations.  

Based on the density distribution, we see that a simple density 
threshold allows to separate them easily. To demonstrate it we have 
plotted the radius vs mass distribution for structures with 
$n_{mean} < 10^5$ cm$^{-3}$ (top-left panel of Fig.~\ref{struct_prop}),
where it is clear that these structures lie in region I. 
 This is also 
confirmed by the distribution of the $\alpha _{vir}$ parameter, which 
is shown for all structures (left-bottom panel of Fig.~\ref{struct_selec}), structures
with mean density larger than $10^5$ cm$^{-3}$ (right-bottom panel of Fig.~\ref{struct_selec})
and mean density smaller than  $10^5$ cm$^{-3}$ (right-middle panel 
of Fig.~\ref{struct_prop}). Most structures with $n_{mean} > 10^5$ cm$^{-3}$
have $\alpha _{vir}$ very close to 1 (note that there is very little mass in structures with $\alpha _{vir} > 1$
in proportion to the one in structure with $\alpha_{vir} \simeq 1$). 
On the contrary the ones 
with $n_{mean} < 10^5$ cm$^{-3}$ have a distribution that is 
broader and are not heavily dominated by an $\alpha _{vir} \simeq 1$ population. 

 Note that the collapsed structures (with high mean density) are very compact and sometimes 
only few cells across. The origin of $\alpha _{vir} \simeq 1$ is the numerical 
diffusion which spreads the density peak few grid cells, while the  typical velocity dispersion that is induced by the 
numerical scheme is simply the virial one. 

In the following 
we will therefore distinguish between objects of 
various mean densities. In our simulations, only the ones with mean densities
below $ \simeq 10^5$ cm$^{-3}$ can be possibly considered as pre or protostellar 
cores. The objects with high mean densities are subject to unphysical merging 
since in practice these objects should have collapsed and formed a star population.  
As will discuss below, their mass distribution is likely affected by this process.

\subsection{Velocity dispersion, Mach number and virial parameter}
The mass vs radius distribution for structures of densities below $10^5$ cm$^{-3}$
is displayed in top-left panel of Fig.~\ref{struct_prop}. It broadly follows an 
$M \propto R^3$ relation with masses on the order of 10 M$_\odot$ for radius of 0.1 pc. 
This is very similar with what has been inferred in the simulations of \citet{offner2008} (see their Figure 1),
more particularly  their undriven case. 
 Note that the trend $M \propto R^{3}$ is likely an artifact of the finite resolution and the 
density threshold of the clump finder. In particular, this relation corresponds to 
the lower mass object at a specific radius. As will be discussed later (section~\ref{thermal_core}), 
the thermally supercritical clumps, that is to say the dense cores, follow a different trend
that is likely not suffering this bias.

The   inner velocity 
dispersion of the objects with $n_{mean} < 10^5$ cm$^{-3}$ is displayed in top-right panel 
of Fig.~\ref{struct_prop}. The distribution is broad, it peaks around 0.5 km s$^{-1}$
but extend for few objects above 1 and below 0.1. Since the sound speed within the dense 
gas is typically on the order of 0.2 km s$^{-1}$, this corresponds to a Mach number  on the 
order of 2-2.5 (not displayed here for conciseness).
There is, as expected a mild
correlation between the mass and the Mach number, $M \propto {\cal \sigma}^{1/2}$ (see the yellow pixels which 
contain most of the mass). 
This is also similar to the values inferred by \citet{offner2008} (their figure 3).

The virial parameter, $\alpha_{vir}$ is displayed in right-middle panel. As can be seen there is, 
as expected a large spread but most of the mass tends to lie in the vicinity of 
$\alpha_{vir} $ on the order of, or slightly  larger than 1. Since real observations
do not have access to $\alpha_{vir}$, we also estimated $\alpha$ using the 
standard definition recalled in Eq.~(\ref{alpha_def}). The two distributions 
are similar without being identical. There is a trend 
toward slightly larger values of $\alpha$. Also its distribution is 
broader than the one of $\alpha_{vir}$.

\subsection{Mass-to-flux ratio and Alfv\'enic Mach number}
The Alfv\'enic Mach number is displayed in left-bottom 
panel of Fig.~\ref{struct_prop}. Typical values are $\simeq$2 times below 
the Mach number ones indicating that the magnetic 
support dominates over the thermal one. Most objects
are sub or trans-Alfv\'enic with very few  values 
larger than 3. There is a clear, though shallow, trend 
for more massive objects to present larger ${\cal M} _{alf}$.
Typically  we get ${\cal M} _{alf} \propto M^{1/4}$. 

The mass-to-flux ratio, $\mu$, is displayed in right-bottom panel. 
Objects for which $\mu$ is below 1 are magnetically subcritical and are not expected to 
undergo gravitational collapse at least as long as they keep their magnetic flux. 
As can be seen there is a clear trend for $\mu$ to increase slightly sub-linearly with the mass
although there is a broad distribution with variation over about one order of magnitude. 
This behaviour is significantly different from studies performed on larger scale clumps
identified through simple density thresholds \citep{banerjee2009,inoue2012,iffrig2017}
where a shallower relation $\mu \propto M^{0.4}$ has been reported. A simple geometrical 
explanation of this relation has been proposed by \citet{iffrig2017}.

The origin of this difference of behaviour between the self-gravitating cores and the diffuse clouds, 
is not obvious. Strictly speaking it 
implies that the magnetic flux is roughly constant through the selected structures or increases 
very mildly with the mass. Since the surface is proportional to $R^2$ and therefore increase 
with the mass, this means that for dense cores, the magnetic field decreases with their mass. 
The most likely explanation, is that matter preferentially flows along the field lines, 
therefore leading a dependence of the mass-to-flux ratio shallower than the one of the large 
scale clumps whose formation  is  primarily due to turbulence.

Another, not exclusive  possibility is that  magnetic diffusion is effective. Indeed magnetic diffusion 
has clearly been observed in the context of collapsing cores \citep{hennebelle2011,joos2013}
although only in the inner part of the cores. Since the dense structures selected by the 
HOP algorithm are local density maximum, there are also regions of the flow which have a 
high magnetic field and since turbulence is significant (being the dominant or comparable to the 
dominant source of support), the clumps experience a few turbulent crossing-times before they collapse.
Note that it cannot be excluded that numerical diffusion is playing an important role in this process
although turbulent diffusion is certainly known to be acting efficiently \citep{lazarian1999}.

The values of $\mu$ indicate that most structures above one solar mass are supercritical 
and vice versa. This certainly suggests that magnetic field plays 
a significant role for the star formation process since it stabilizes most of the small 
clumps that form, a point that we will discuss further in the following.
It should also be stressed that while the value of $\mu$ are typically larger than 1 for 
massive cores, most of them are still below 10 which indicates that the magnetic field 
still has a significant influence during the collapse \citep[e.g.][]{hennebelle2011,commercon2011,myers2013}.
In particular, magnetic fields of such intensities can play an active role in reducing the 
gravitational fragmentation that may occur during collapse.

\subsection{Mass spectra}
An important statistical property regarding the prestellar cores is 
their mass spectrum. Indeed it has been found that 
the core mass spectrum is very similar in shape to the IMF
 \citep{motte98,alves2007,andre2010,konyves2015} and 
several theories have been assuming that the core mass function (CMF)
is at the origin of the IMF \citep{padoan1997,hc2008,hopkins2012,offner2014}. 
While the link between the CMF and the IMF is still debated, the CMF 
provide an important statistical description of the dense and self-gravitating gas, 
that needs to be reproduced and understood.

Figure~\ref{mass_spectra} shows several mass spectra of various ensemble 
of structures. Top panel displays the mass spectrum of all structures 
identified by the HOP algorithm in the simulation and having at least 100 computing cells.
The solid lines indicate for reference the mass spectra $dN / d \log M \propto M^{-1}$ and $dN / d \log M \propto M^{-1.3}$. 
The mass spectrum of all structures ranges from masses of about 0.01 $M_\odot$ to masses larger than 10$^3$  $M_\odot$. 
The high mass part (above 10  $M_\odot$) presents a clear  $M^{-1}$ tail. 
The low mass part peaks at about 0.1 $M_\odot$ and then it steeply drops. 

As seen from Figs.~\ref{struct_selec} and \ref{struct_prop}, many structures are not gravitationally bound 
or have already collapsed and should not be considered as prestellar cores. Therefore 
we also show the mass spectra of various sub-populations. The left-middle panel shows 
the mass spectrum of structures that have a mass-to-flux ratio, $\mu$, smaller than 1, that 
is to say subcritical structures while the right middle panel shows the mass spectrum of super-critical 
cores (black lines) and cores having $\mu>0.3$. 
Clearly the $\mu$ parameter controls the peak of the magnetized core distribution. The subcritical structures 
present a peak at about $0.2 \, M_{\odot}$ and does not present  a power-law
distribution at high mass. 
Instead its shape is roughly lognormal.  We caution that, as already discussed, the definition and 
therefore the physical meaning of many subcritical clumps, should be regarded with care. 
In particular the peak depends on the numerical resolution.
On the contrary supercritical
cores (with $\mu > 1$) have a mass spectrum which peaks at about $2 \, M_{\odot}$ and 
present a high mass tail $\propto M^{-1}$. Unsurprisingly the peak shifts 
toward smaller mass for larger values of $\mu$. 

To remove the collapsed cores discussed in the previous section, we have selected 
supercritical cores for which $n_{mean} < 10^5$ cm$^{-3}$ (left-bottom panel, black line)
and $n_{mean} < 10^6$ cm$^{-3}$ (left-bottom panel, blue line). 
 The low mass part is nearly identical to the supercritical 
core mass spectrum displayed in the right-middle panel but the high mass tail is 
quite different. It is still a power-law but much closer to be 
$\propto M^{-1.3}$ than  $\propto M^{-1}$.

Finally, we have also plotted the mass spectra for thermally supercritical cores,
 that is to say for which $\alpha_{th}<1$ keeping 
again the ones for which $n_{mean} < 10^5$ cm$^{-3}$ (black line of 
right-bottom panel) and $n_{mean} < 10^6$ cm$^{-3}$ (blue line).
The motivation is twofold. First of all as already mentioned, ambipolar diffusion is not included
and could reduce the magnetic flux,
second of all, observationally it is hard to measure the magnetic intensity
and for this reason thermal support is usually considered to select 
gravitationally bound cores. As can be seen, the shape of the high mass part is 
identical to the supercritical cores. 
Both  mass spectra  peak at about $1-2 \, M_{\odot}$. 
There are however more small cores in the thermally supercritical distribution 
than in the magnetically supercritical one and the former is slightly 
broader than the latter.

Altogether these results are reminiscent of the core mass functions, that 
have been observationally obtained. In particular \citet{andre2010} 
found that in the Gould Belt survey, the CMF peaks around or slightly below 
$1 \, M_{\odot}$ and present a power-law  $\propto M^{-1.3}$ at high mass. 
On the contrary the mass spectrum of the structures observed in the Polaris cloud, which
are not self-gravitating,  peaks
at smaller mass and has a lognormal shape. This is reminiscent of the 
mass spectra obtained here. The mass spectrum of subcritical structures (left-middle panel) resembles the Polaris
one and the mass spectrum of supercritical ones (left-bottom panel) are similar 
to the CMF obtained for the Gould Belt although the observational CMF 
may peak at a value $\simeq 2-3$ smaller than the one inferred from the simulation (but see
section \ref{convergence} for a discussion on possible numerical convergence issue).

Our results are also reminiscent of some of the core mass functions 
previously obtained in numerical simulations \citep{klessen1998,klessen2001,gong2015}
that also present a peak and powerlaws at high masses. We stress however that since these studies
are isothermal, the core masses can be freely  normalized. In the present simulation, cooling is treated 
and more generally the density distribution is a consequence of several processes, such as the 
disc vertical equilibrium itself related to the momentum injected by the supernovae.

While this is encouraging, it is important to stress that there may be difficult issues however 
regarding the numerical resolution and the dependence of the peak position with it, something 
that we discuss in Section~\ref{convergence}.

\subsection{Properties of thermally supercritical cores}
As our main interest are  the supercritical cores, we now specifically investigate 
some of their properties. 

\subsubsection{Mass-radius of thermally supercritical cores}
\label{thermal_core}

Upper panel of Fig.~\ref{mu-th-vir} shows the mass-radius relation for the thermally 
supercritical cores. Apart for the very low mass ones, the distribution is 
broadly encompassed between  $M \propto R$ for the most massive objects at a specific radius and 
 $M \propto R^2$ for the less massive ones
though the dispersion is quite large  for $\log R $ below -1.5. 
The overall distribution is broadly similar with the one inferred by 
\citet{konyves2015} (see their Fig.~7).

\subsubsection{Mass-to-flux ratio of thermally supercritical cores}
We now examine the correspondence between the thermally and magnetically supercritical cores. 
For that purpose we study the distribution of the mass-to-flux ratio, $\mu$, for 
cores having $\alpha_{th} < 1$ and mean density below $10^5$ cm$^{-3}$. 
Lower panel of Fig.~\ref{mu-th-vir} displays the result. As can be seen while most massive cores are clearly
magnetically supercritical (i.e. have $\mu >1$), this is less the case for low and intermediate 
mass cores for which a significant fraction are actually dominated by magnetic field. 
While this result was expected since the mass spectrum of thermally supercritical cores is broader 
than the mass spectrum of magnetically supercritical one, 
this nevertheless illustrates the difficulty of defining what a core exactly is. Indeed, a thermally
subcritical object may accrete more mass or be compressed and this could make 
it gravitationally unstable. Similarly magnetically subcritical cores may accrete mass along the field lines or lost some 
magnetic flux through ambipolar diffusion if it is held by external pressure for a few diffusion times. 

Note we have not included kinetic energy in the core selection, which may make some of the 
thermally supercritical cores stable. The reason is that in this process, one should carefully distinguish 
between collapsing motion that should not be counted as kinematic support but rather counted negatively. 
This would require careful analysis that goes beyond the present paper. Qualitatively, the mass spectrum 
is similar but the peak position is even more uncertain.  

Observationally, \citet{crutcher2012} inferred that most cores are supercritical while only 
a few appears to be subcritical. While it may simply be an effect of selected samples (most of the selected 
observed cores may have already formed an object or are on the verge to form one while our ``subcritical'' cores
simply expand without forming an object), 
this may also possibly indicate that either ambipolar diffusion 
should be included as it is playing a significant role at the scale on the order of 
0.1 pc, either magnetic field is a bit too high in the present simulations.

\subsubsection{Angular momentum}
Angular momentum is an essential  quantity in the context of core collapse and  disc formation
and we therefore investigate its distribution in our core population. 
The specific momentum is given by 
\begin{eqnarray}
J =   { \| \int ({\bf v- v_0}) \times {\bf r} \,  dm \|  \over \int dm }.
\end{eqnarray}

Figure~\ref{ang_mom} shows its distribution for supercritical cores with mean density below $10^6$ cm$^{-3}$. 
Upper panel displays its value as a function of mass while bottom one shows it as a function of 
the velocity dispersion. As can be seen the inferred values go from $10^{-3}$ to $10^{-1}$ 
pc km s$^{-1}$ and scales with the mass roughly as $M^{2/3}$ implying that $J \propto R^2$. These values are in excellent agreement 
with what has been inferred from observations (see for example Fig.~7 of \citet{belloche2013}). 
It is also compatible with previous simulations such as the ones 
performed by \citet{offner2008} (their figure 5) and \citet{dib2010}  (their figure 13). 

Interestingly, the correlation between $J$ and $\sigma$ is slightly better than between $J$ and $M$. 
This is in good agreement with the idea that the rotation of pre and proto-stellar 
cores is primarily inherited from their initial turbulence.

\setlength{\unitlength}{1cm}
\begin{figure*} 
\begin{picture} (0,18.5)
\put(0,12){\includegraphics[width=8.7cm]{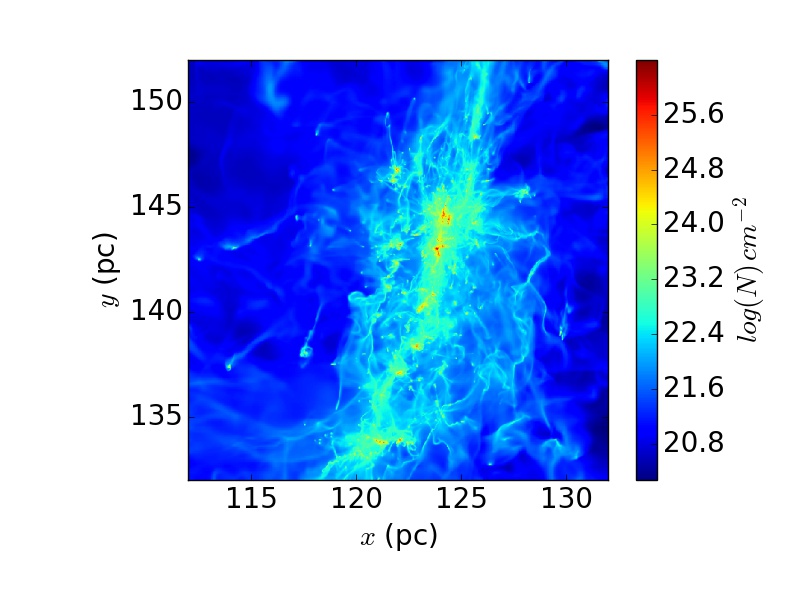}}  
\put(9,12){\includegraphics[width=8.7cm]{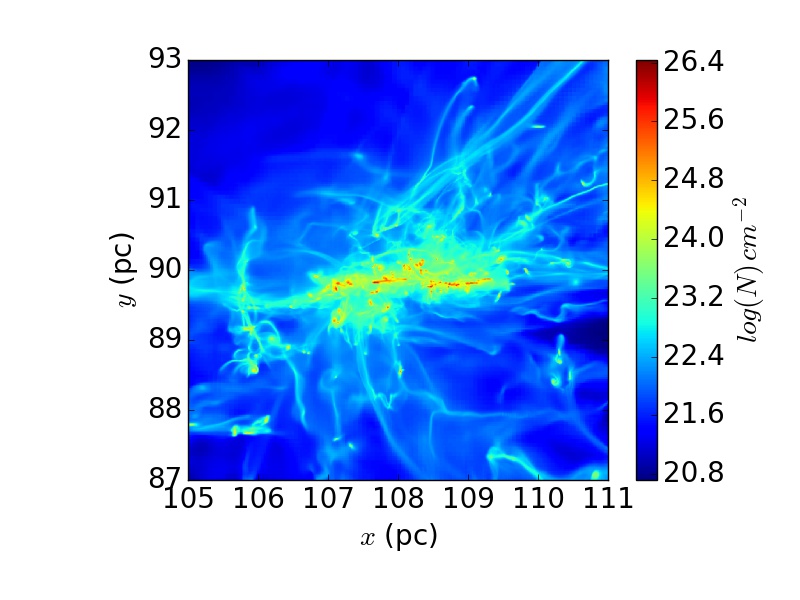}}  
\put(0,6){\includegraphics[width=8.7cm]{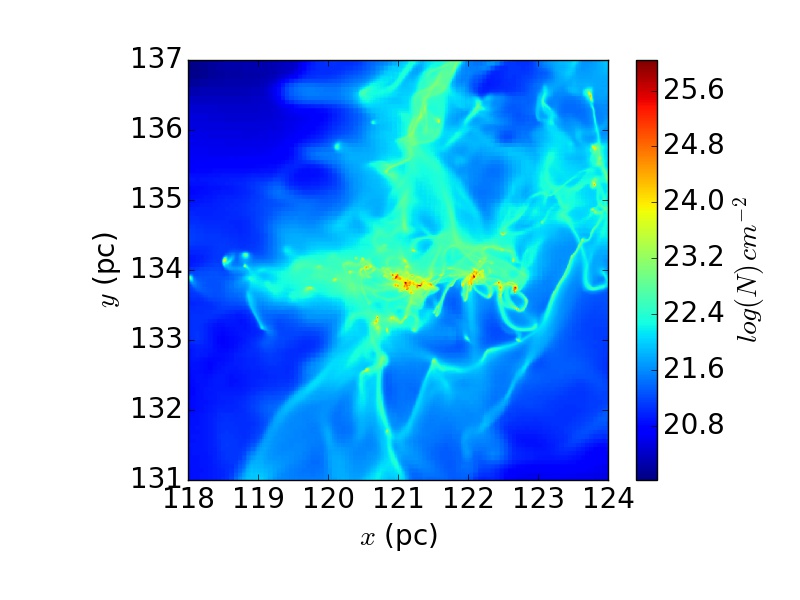}}  
\put(9,6){\includegraphics[width=8.7cm]{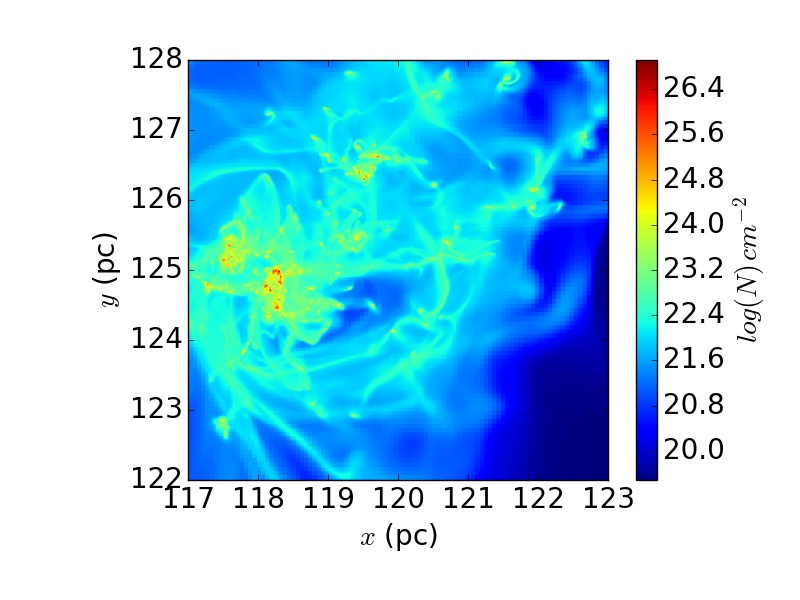}}  
\put(0,0){\includegraphics[width=8.7cm]{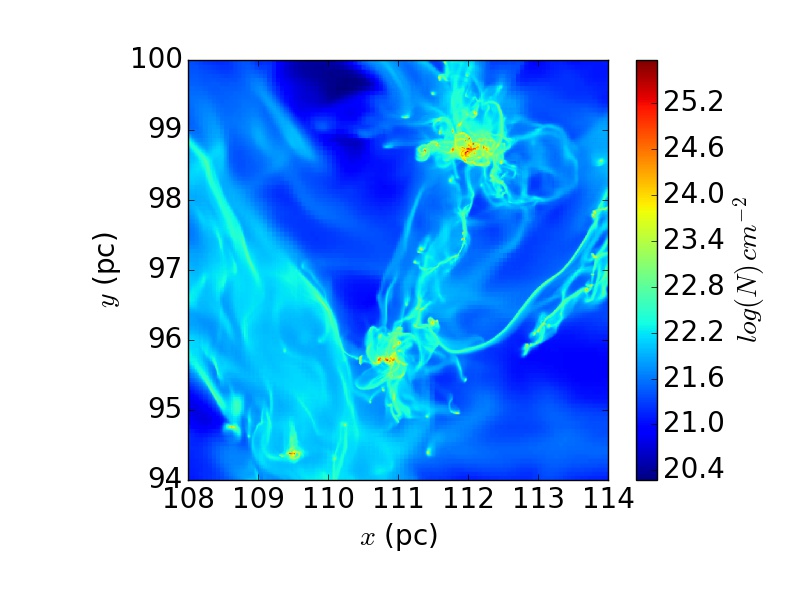}}  
\end{picture}
\caption{Column density for the five selected sub-regions.}
\label{clust_im}
\end{figure*}

\setlength{\unitlength}{1cm}
\begin{figure} 
\begin{picture} (0,17.5)
\put(0,5.5){\includegraphics[width=8cm]{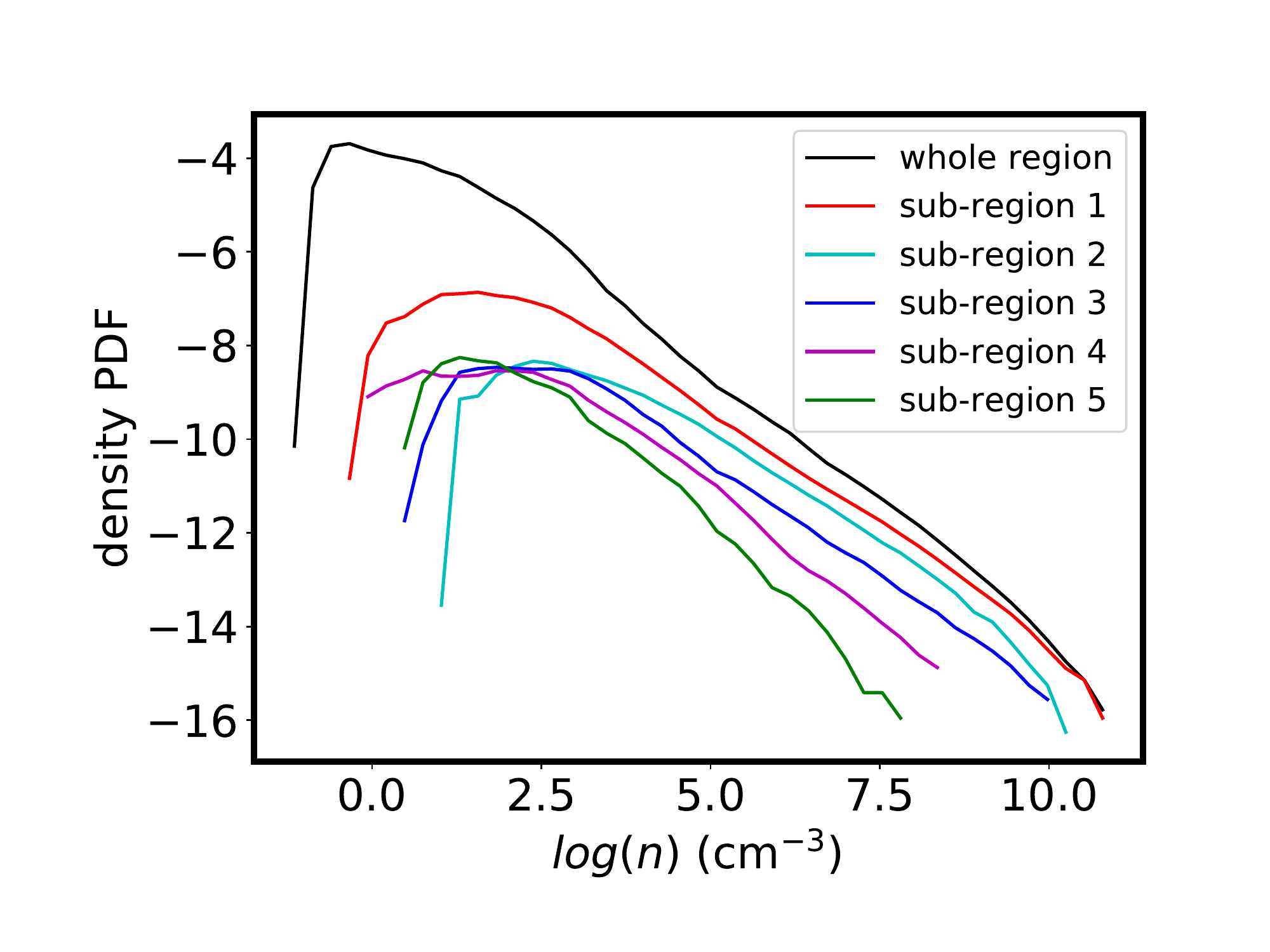}}  
\put(0,11){\includegraphics[width=8cm]{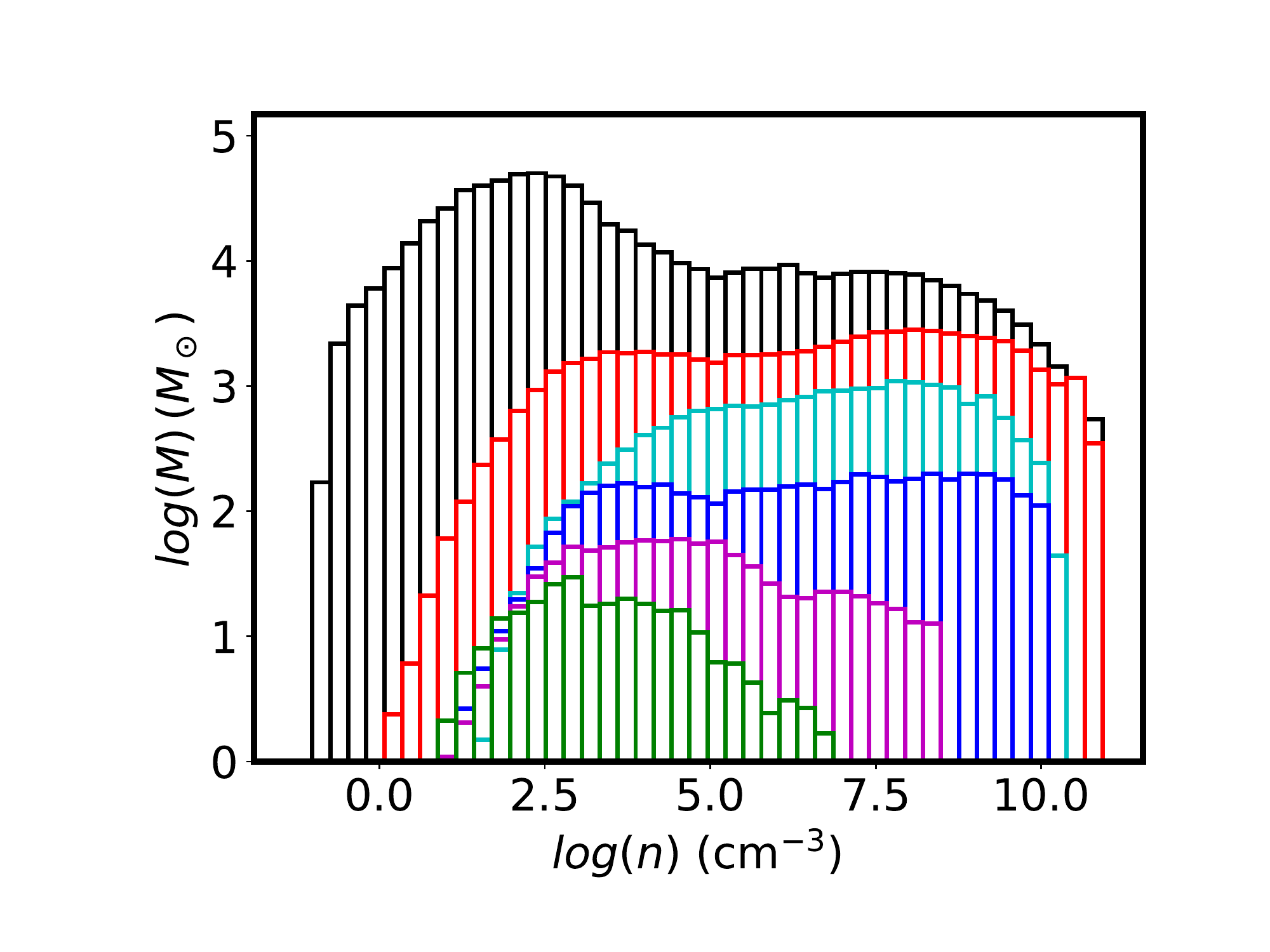}}  
\put(0,0){\includegraphics[width=8cm]{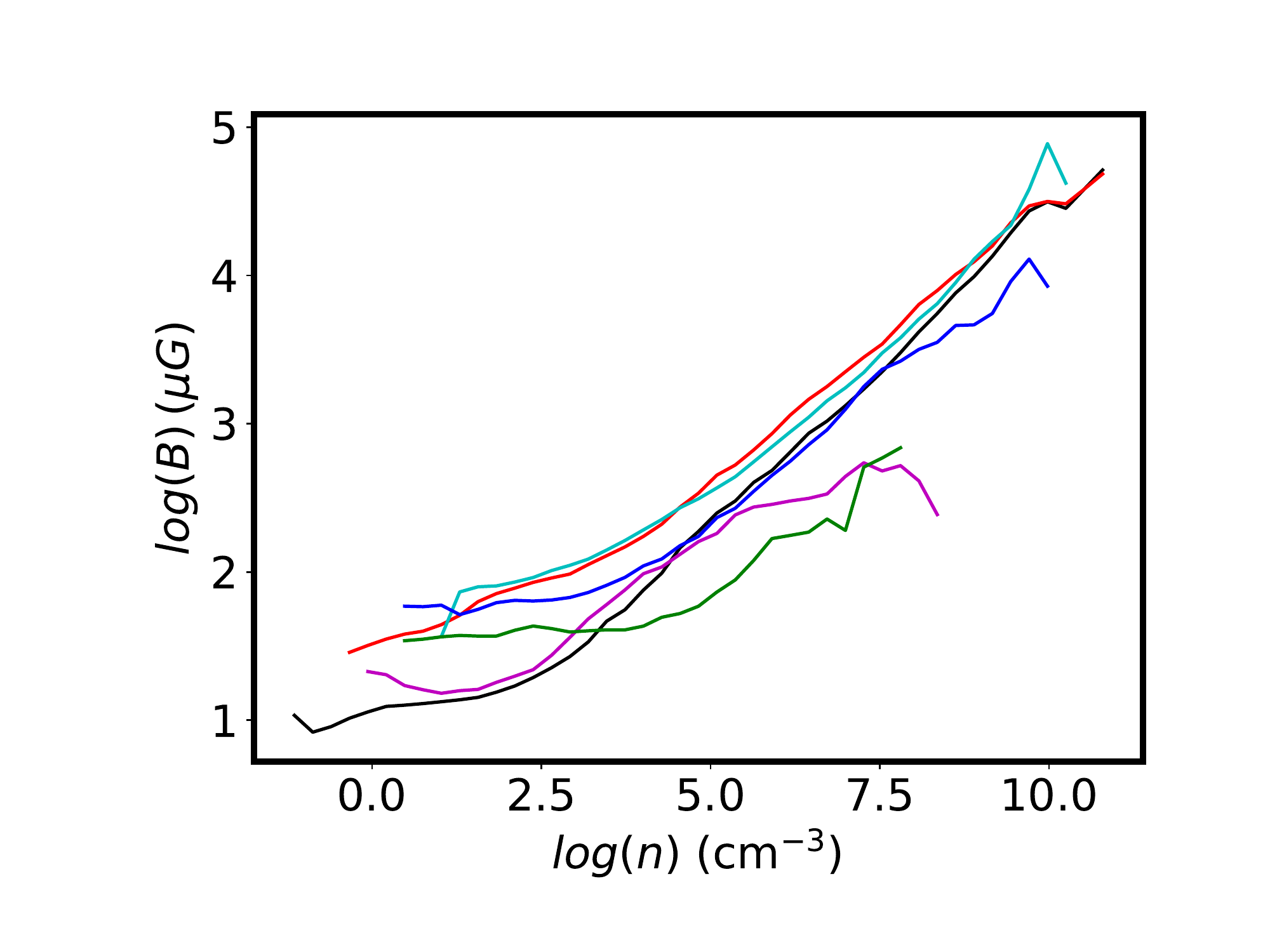}}  
\end{picture}
\caption{Top panel: mass distribution as a function of density (mass weighted density PDF), 
bottom panel: mean magnetic intensity as a function of density for the five sub-regions 
shown in Fig.~\ref{clust_im}.
 }
\label{pdf_clust}
\end{figure}

\setlength{\unitlength}{1cm}
\begin{figure} 
\begin{picture} (0,11.5)
\put(0,5.5){\includegraphics[width=8cm]{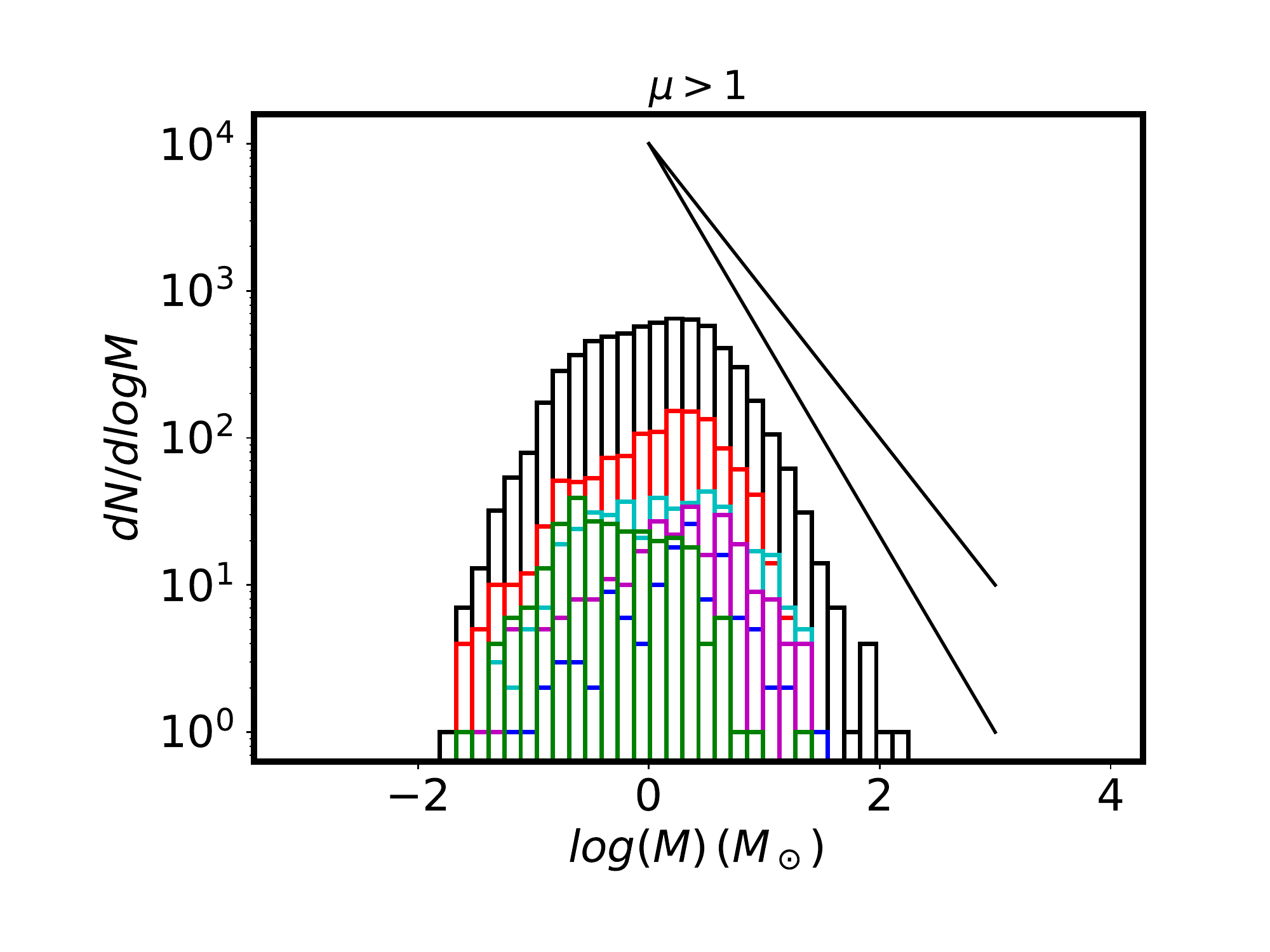}}  
\put(0,0){\includegraphics[width=8cm]{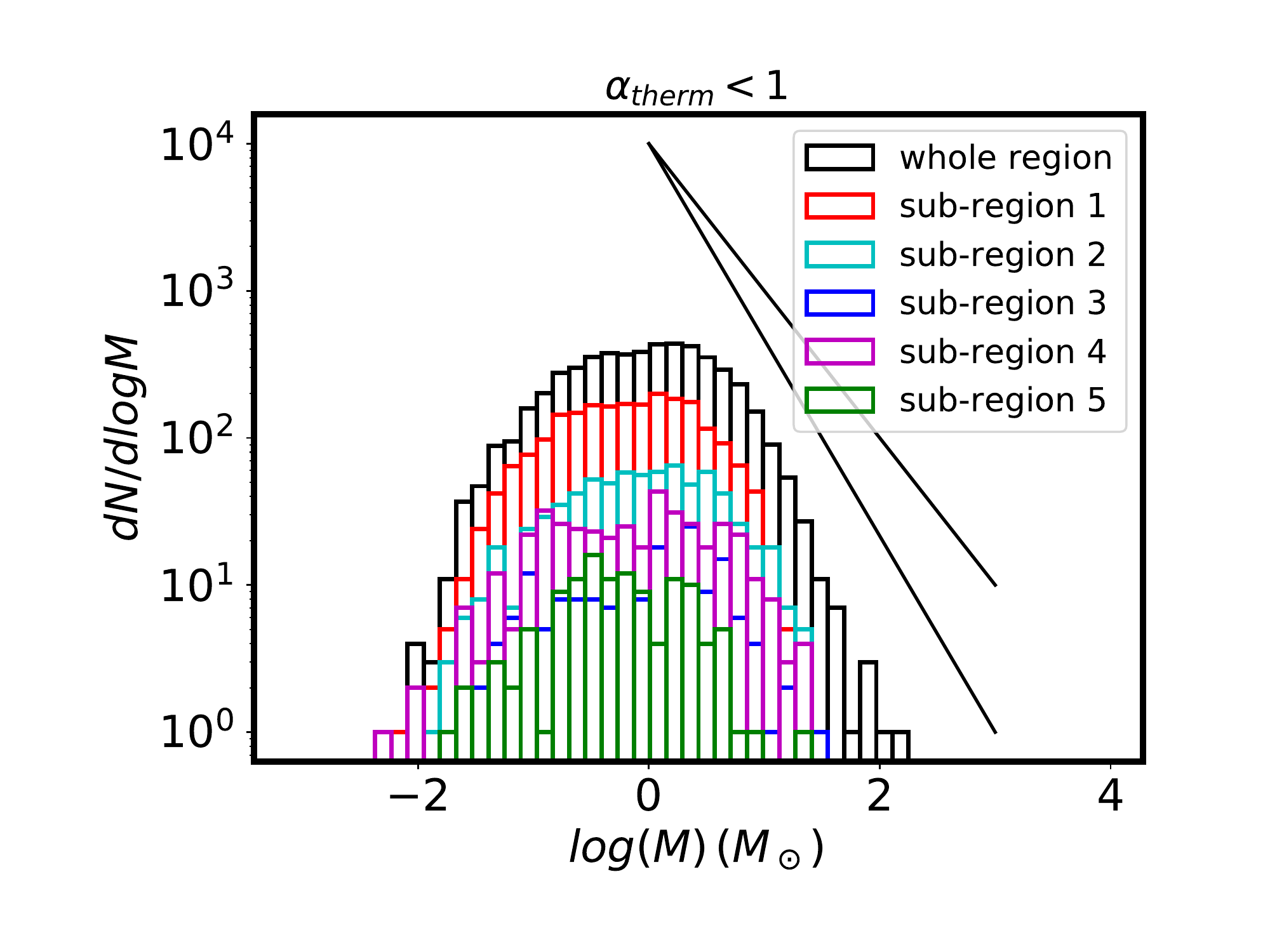}}  
\end{picture}
\caption{Core mass function at time $t=10$ Myr for the Z18 simulations and for 
five sub-regions.
 }
\label{mass_spectra_clust}
\end{figure}

\section{Environmental dependences}
So far we have considered the statistics for all the extracted cores 
present in the simulation. An important question is to what extent the
core properties may vary from regions to regions. Given the 
relatively large simulated volume, there is indeed a wide range 
of physical conditions. To tackle this question, we have selected five 
subregions of the simulation Z18 at time 10.04 Myr as displayed
in Fig.~\ref{clust_im}. These regions contain many cores, therefore statistics 
can be drawn. 

\subsection{Physical characteristics}
First we quantify key physical characteristics of the 
selected sub-regions. Figure~\ref{pdf_clust} displays 
the density PDF, the mass distribution or equivalently the 
mass-weighted density PDF, and the magnetic intensity 
as a function of density in the five sub-regions as well as 
in the whole fully refined zoom region.

As displayed by the top panel, the five regions present 
very different masses going from few 10$^4$ M$_\odot$ (region 1)
to about 100 M$_\odot$ (region 5). They present density 
PDF (middle panels) that have similar shape.
They peak at low densities around 10 to 100 cm$^{-3}$ 
(1 cm$^{-3}$ for the whole refined region) and a powerlaw  
  $\simeq \rho^{-1}-\rho^{-1.5}$ at high density. 
Similar distributions have been found to be 
typical of gravitational collapse
\citep{kritsuk11,Hennebelle12,girichidis2014}. The
 less massive subregions (5) presents however 
significant deviations at high densities possibly indicating that it contains 
less collapsed objects.  

The mean magnetic intensity is displayed as a function of density in 
bottom panel. 
For the whole region, the usual behaviour \citep{troland86,hennebelle2008,banerjee2009,crutcher2012} 
is recovered, that is to say 
$B$ weakly depends on $n$ for $n < 10^3$ cm$^{-3}$, where typical magnetic 
intensities are $\simeq 10$ $\mu$G, while at densities $n > 10^4$ cm$^{-3}$,
$B \propto \sqrt{n}$. Let us remind that this behaviour is a simple 
consequence of the magnetic and gravitational forces. In the diffuse gas, gravity is not dominant 
and the gas must flow along the field lines to avoid high magnetic pressure. On the contrary in the dense gas, 
gravity can compensate for the high magnetic pressure \citep{hennebelle2008}. 
For the subregions 1, 2 and 3 a similar behaviour is inferred although 
regions 2 and 3 present values of $B$ at low densities that are 3-10 times higher.
This is due to the fact that the magnetic field has been globally compressed by gravity 
in these regions. Subregions 4 and 5 present a slightly different behaviour at high densities
particularly region 5, that presents magnetic intensity values 2-3 times below the others. 

We conclude that while the five  subregions present rather different masses, their physical conditions 
at more similar except subregion 5, which present lower magnetic intensities.

\subsection{Mass spectra of the sub-regions}

Figure~\ref{mass_spectra_clust} shows the CMF for the whole region and the five 
subregions. Top panel shows the magnetically supercritical cores while bottom one
displays the thermally supercritical cores. 

Subregions 1 and 2 present mass spectra that are relatively similar to the whole region one.
The peak is approximately at the same place and the shape of high-mass part of the distribution 
also resembles the CMF of the whole region.

Subregion 3 presents more fluctuations, which is expected since it contains less cores.
Its CMF is nevertheless similar to the ones of regions 1 and 2. 

Subregions 4 and 5 present more systematic deviations. 
There is an excess 
of low mass  magnetically supercritical  cores (top panel) of subregion
5 (which peaks at about 0.3 M$_\odot$) as well as a paucity of high mass cores. 
A similar behaviour is  observed for thermally supercritical cores of region 5.
This is entirely consistent with the magnetic intensity distribution discussed above. 
Subregion 4 also shows an excess of thermally supercritical cores but not of  magnetically 
supercritical cores.

Altogether these results suggest that the core properties  (note that only the  mass distribution
has been displayed here for conciseness but similar results are obtained for the other ones) does not 
fluctuate very strongly from one region to another provide the region is massive enough.  
Note that this lack of strong variations is particularly important in the context of the apparent universality of 
the IMF \citep[e.g.][]{bastian2010} since cores are believed to be 
the mass reservoir of stars although the links between the IMF and the CMF is still debated. 
At first sight these weak variations are not straightforward to understand because there is a broad range of 
density, velocity dispersion and magnetic field in the zoomed region. 
We believe that the limited variations of the CMF may be a consequence 
of the virial dynamical equilibrium that naturally develops in a collapsing turbulent clump, 
and tends to {\it self-regulate the initial conditions} of star forming clumps
 as proposed by \citet{lee2016a,lee2016b}. In particular, the effects of the density and velocity 
variations onto the CMF tend to compensate each others within  such  regions \citep{hennebelle2012,lee2016b}. 

\section{Time evolution and  numerical convergence}
\label{converg}

In this section, we discuss the robustness of the results presented above. 
Indeed the zooming strategy may introduce biases such as a dependence 
on the time at which zooming starts or on the zooming strategy itself. Here we investigate 
in details the evolution of the statistics with time as well as the impact of 
the maximum resolution reached in  the simulations onto the statistics.

\setlength{\unitlength}{1cm}
\begin{figure} 
\begin{picture} (0,11.5)
\put(0,5.5){\includegraphics[width=8cm]{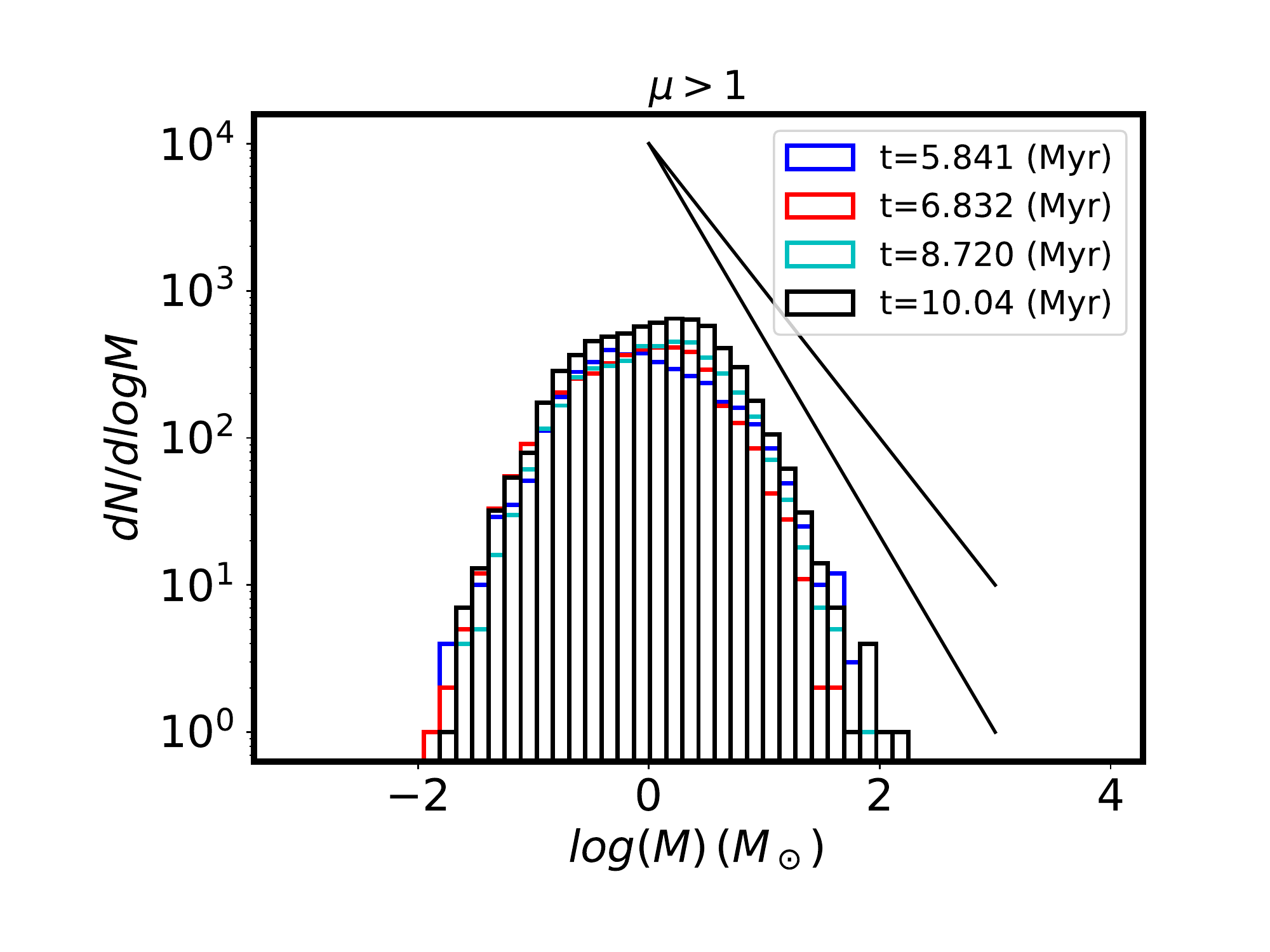}}  
\put(0,0){\includegraphics[width=8cm]{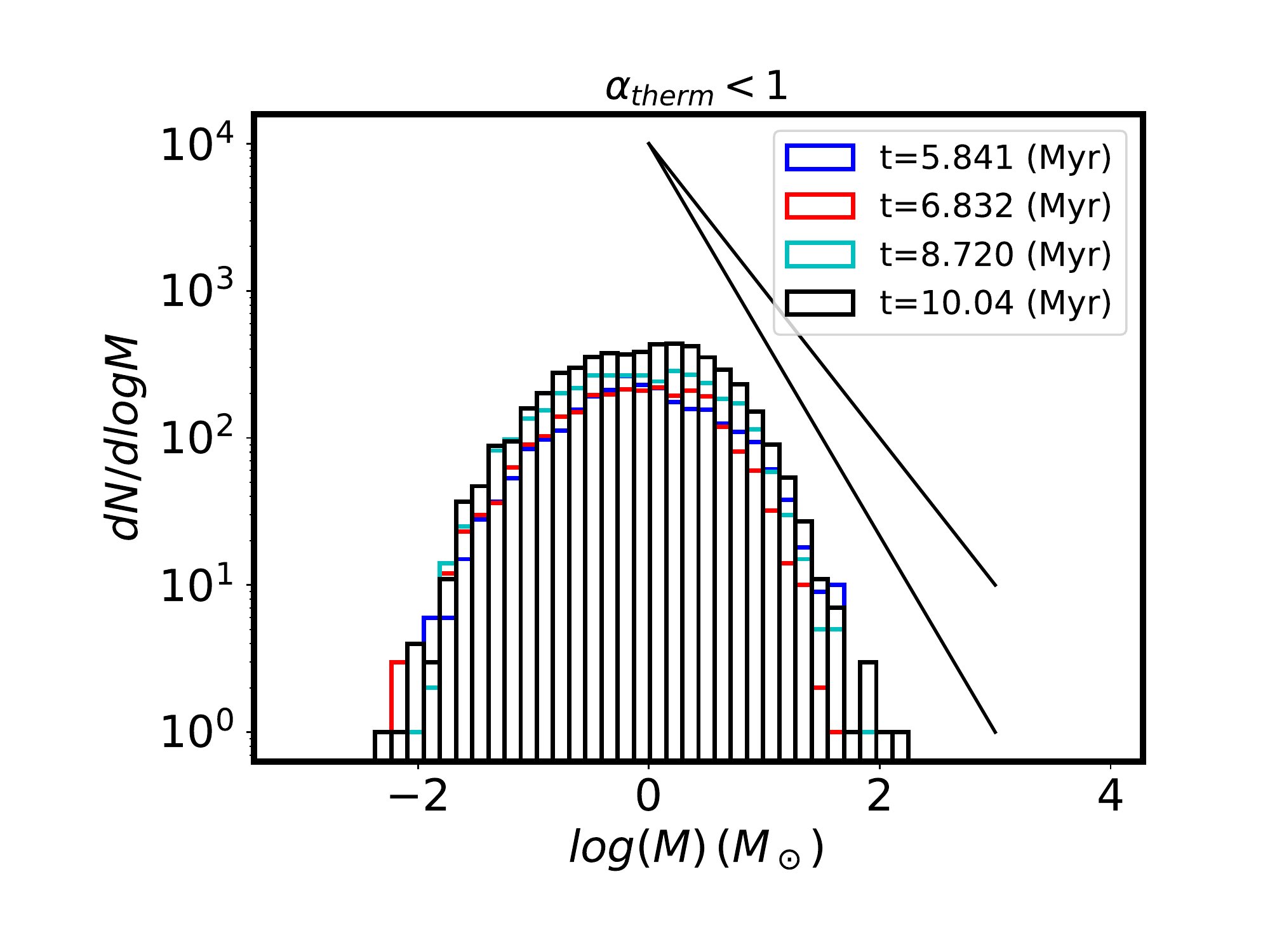}}  
\end{picture}
\caption{Core mass function for simulation Z18 at different timesteps. 
 }
\label{mass_spectra_evol}
\end{figure}

\setlength{\unitlength}{1cm}
\begin{figure} 
\begin{picture} (0,11.5)
\put(0,5.5){\includegraphics[width=8cm]{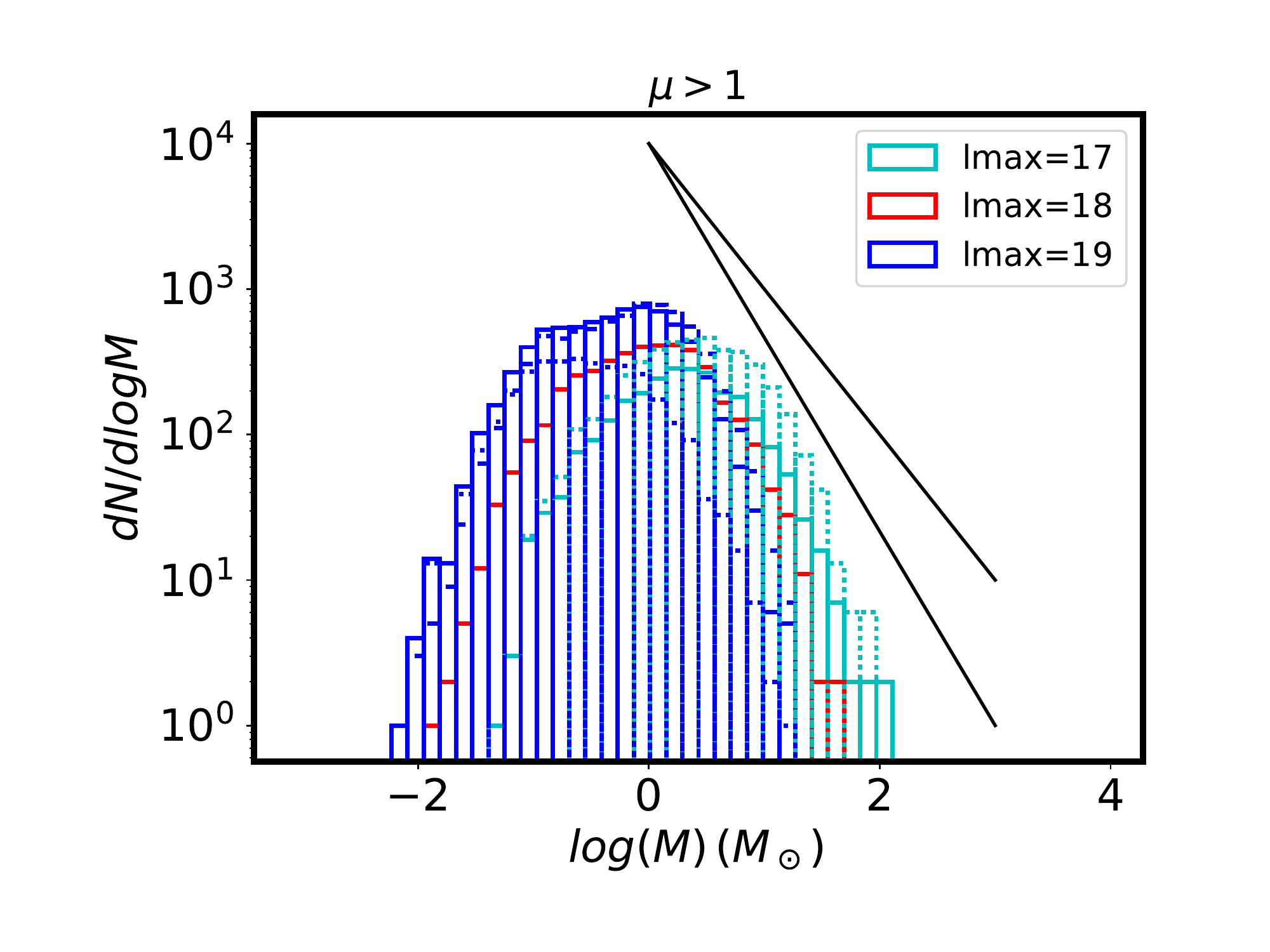}}  
\put(0,0){\includegraphics[width=8cm]{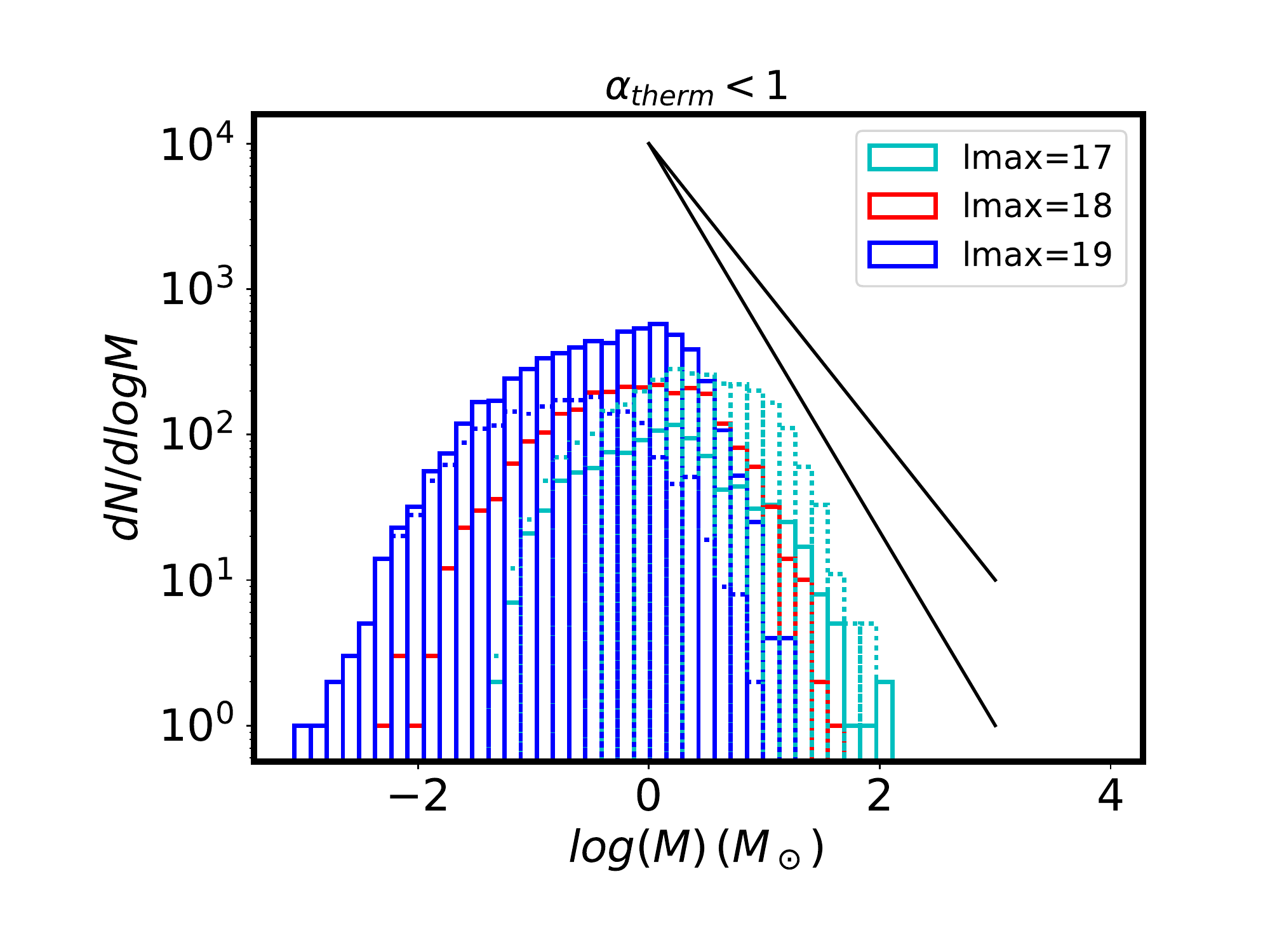}}  
\end{picture}
\caption{Core mass function for three numerical resolutions, simulations Z17, Z18 and Z19.
 }
\label{mass_spectra_res}
\end{figure}

\subsection{Time evolution}
\label{time_evol}
 
Figure~\ref{mass_spectra_evol} shows the CMF (top panel  uses the $\mu$ parameter
while bottom one uses $\alpha_{therm}$, for both of them $n_{mean} < 10^5$ cm$^{-3}$) for four timesteps of the Z18 simulation.
The difference between the first and the last timestep
presented is about 4 Myr, which represents about one freefall time for a gas  density 
of  100 cm$^{-3}$. This therefore implies that gas typically denser than 10$^3$ cm$^{-3}$
at time 5.841 Myr will have experienced about 4 dynamical times at time 10.04 Myr. Thus 
the early cores present  at time 5.841 Myr have collapsed at time 10.04 Myr, while the 
early cores present at time 10.04 Myr are made out of gas that was diffuse enough at time 
5.841 Myr. By comparing the CMF at these two times, and more generally the CMF evolution, 
we can therefore investigate to what extent the CMF we infer depend on the zooming-time and 
on the zooming strategy.  

The first time step ($t=5.841$ Myr) is only 0.2 Myr after the full refinement starts, that is to say 
refinement based of Jeans length  up to level $l=18$. As can be seen there is a 
visible evolution between time 5.841 and 6.832 Myr for small objects. In particular the 
peak for the CMF based on $\mu$ is shifting by a factor $\simeq 2$. At later time, 
we see that the number of objects increases by a factor of the order of 2 but that the shape of the distribution 
does not evolve significantly. Since these objects form a few millions years after 
the full resolution starts, this clearly indicates that the  starting time does not 
drastically affect the statistics of the cores. 

The zooming strategy is likely responsible of the difference between $t=5.841$ and $t=6.832$ Myr. 
However since there is no strong evolution of the shape at later time, the zooming strategy 
we used seem to lead  quickly to result equivalent to the ones obtained with full refinement.

\subsection{The issue of numerical convergence}
\label{convergence}

Finally, we investigate the influence of the numerical convergence, which is a central issue 
for the CMF. Figure~\ref{mass_spectra_res} shows the CMF for Z17, Z18 and Z19 at comparable times,
i.e. 0.8 Myr after the beginning of full refinement, which was the latest we could 
achieve for the Z19 run. There is however a difficult point here. As discussed previously 
the mean density threshold is playing an important role to remove the  collapsed
objects. However, resolution obviously affects the mean density  of collapsed objects, 
which increases as $dx^{-3}$, where $dx$ is the finest spatial resolution. This implies
that the threshold density should increase by  a factor 8 when the maximum resolution 
improves by a factor 2 (for example from level 18 to level 19).
Therefore 2 thresholds are used for Z17 and Z19 runs. The dashed lines 
 show the CMF for cores with mean density below 10$^5$ cm$^{-3}$ while the solid lines
show the CMF with a threshold of 10$^6$ cm$^{-3}$ for Z19, 10$^5$ cm$^{-3}$ for Z18
and 10$^4$ cm$^{-3}$ for Z17. 

As can be seen the CMF of the Z17, Z18 and Z19 simulations with a mean density threshold 
of 10$^5$ cm$^{-3}$, are quite different. They peak at 5, 1 and 0.2 $M_\odot$ respectively. 
This is expected since smaller structures are more numerous when the resolution is higher. 
However the CMF of the Z17, Z18 ad Z19 simulations with threshold respectively 
equal to 10$^4$, 10$^5$ and 10$^6$ cm$^{-3}$ are much closer and peak respectively 
at about 2, 1.5 and 1 $M_\odot$. Interestingly the high mass part of the 
3 CMF are also close with differences of about a factor 2. The differences are more 
pronounced for the thermally supercritical cores. 

It is thus not possible to conclude at this stage, that numerical convergence has been reached. 
Clearly many cores have pronounced internal structure  and defining unambiguously what is a core 
may be an elusive task.   
The question as to whether the CMF will eventually converge is therefore not straightforward
although \citet{gong2015} seems to get numerical convergence in their 
simulations (see their figure 10). It may 
eventually depend on the small scale processes that  determine whether they fragment or not.

\section{Conclusion}

We have carried out zooming simulations using self-consistently generated 
initial conditions from  a self-regulated supernova ISM at the kpc scale.
Our strategy consists in performing a series of concentric uniform refinement and then 
for the last levels to use the Jeans length as a refinement criterion. 
The spatial resolution goes up to a few $10^{-3}$ pc which is enough to describe 
the formation of cores of masses on the order of few 0.1 $M_\odot$. 
We identify the cores using a clump finder and requiring that the structures 
are thermally or magnetically supercritical. 
Since 
the domain where full refinement is applied extent over 100 pc, we get a few
thousands of cores providing reliable statistics. 
The inferred CMF present clear similarities with the observed ones. 
In particular the massive objects present a powerlaw with an exponent close to 
$-1.3$ similar to the IMF. The peak of the CMF is found to be located 
around 1-2 $M_\odot$, also similar to the observations (though possibly higher by a factor 2-3). Its position 
may however vary with the resolution and the significance of this peak 
must therefore be confirmed by future studies. Other statistics such as the 
velocity dispersion, the angular momentum and the magnetization also 
present encouraging agreements. For instance as in the observation the velocity dispersion 
in cores are typically sonic or mildly supersonic. The angular momentum increases 
with the core radius and typical values are on the order of $10^{-2}$ pc km s$^{-1}$.
The magnetization is significant most cores having a mass-to-flux ratio in the range 0.3 to 3.  
The statistics do not vary significantly 
with time, seemingly suggesting that the zooming procedure used in this paper does not introduce 
severe biases.  \\ \\

\emph{Acknowledgments}
I thank the anonymous referee for a report that helped to 
improve the manuscript. I thank Philippe Andr\'e for a critical reading of the manuscript as well
as Juan Soler, Eva Ntormousi, Sam Geen and Damien Chapon for related discussions.  
This work was granted access to HPC resources of CINES under the 
allocation  x2014047023 made by GENCI (Grand Equipement National de Calcul Intensif).
This research has received funding from the European Research Council under the European
 Community's Seventh Framework Programme (FP7/2007-2013 Grant Agreement no. 306483).
We acknowledge PRACE for awarding us access to resource CURIE based in France at TGCC.

\bibliography{refs}{}
\bibliographystyle{aa}

\end{document}